\documentclass[onecolumn,12pt]{article}
\usepackage{graphicx}
\usepackage{color}
\newcommand\simlt{\hspace{0.3em}\raisebox{0.4ex}{$<$}\hspace{-0.75em}\raisebox{-.7ex}{$\sim$}\hspace{0.3em}} 
\newcommand\simgt{\hspace{0.3em}\raisebox{0.4ex}{$>$}\hspace{-0.75em}\raisebox{-.7ex}{$\sim$}\hspace{0.3em}}

\begin{document}
\baselineskip 14pt

\title{Distortion of Magnetic Fields in Barnard 335} 
\date{Ver.3}
\author{Ryo Kandori$^{1}$, Masao Saito$^{2}$, Motohide Tamura$^{1,2,3}$, Kohji Tomisaka$^{2}$, \\
Tomoaki Matsumoto$^4$, Ryo Tazaki$^{5}$, Tetsuya Nagata$^{6}$, Nobuhiko Kusakabe$^{1}$, \\
Yasushi Nakajima$^{7}$, Jungmi Kwon$^{8}$, Takahiro Nagayama$^{9}$, and Ken'ichi Tatematsu$^{2}$\\
{\small 1. Astrobiology Center of NINS, 2-21-1, Osawa, Mitaka, Tokyo 181-8588, Japan}\\
{\small 2. National Astronomical Observatory of Japan, 2-21-1 Osawa, Mitaka, Tokyo 181-8588, Japan}\\
{\small 3. Department of Astronomy, The University of Tokyo, 7-3-1, Hongo, Bunkyo-ku, Tokyo, 113-0033, Japan}\\
{\small 4. Faculty of Sustainability Studies, Hosei University, Fujimi, Chiyoda-ku, Tokyo 102-8160}\\
{\small 5. Astronomical Institute, Graduate School of Science Tohoku University,}\\
{\small 6-3 Aramaki, Aoba-ku, Sendai 980-8578, Japan}\\
{\small 6. Kyoto University, Kitashirakawa-Oiwake-cho, Sakyo-ku, Kyoto 606-8502, Japan}\\
{\small 7. Hitotsubashi University, 2-1 Naka, Kunitachi, Tokyo 186-8601, Japan}\\
{\small 8. Institute of Space and Astronautical Science, Japan Aerospace Exploration Agency,}\\
{\small 3-1-1 Yoshinodai, Chuo-ku, Sagamihara, Kanagawa 252-5210, Japan}\\
{\small 9. Kagoshima University, 1-21-35 Korimoto, Kagoshima 890-0065, Japan}\\
{\small e-mail: r.kandori@nao.ac.jp}}
\maketitle

\begin{abstract}
In this study, the detailed magnetic field structure of the dense protostellar core Barnard 335 (B335) was revealed based on near-infrared polarimetric observations of background stars to measure dichroically polarized light produced by magnetically aligned dust grains in the core. Magnetic fields pervading B335 were mapped using 24 stars after subtracting unrelated ambient polarization components, for the first time revealing that they have an axisymmetrically distorted hourglass-shaped structure toward the protostellar core. On the basis of simple two- and three-dimensional magnetic field modeling, magnetic inclination angles in the plane-of-sky and line-of-sight directions were determined to be $90^{\circ} \pm 7^{\circ}$ and $50^{\circ} \pm 10^{\circ}$, respectively. The total magnetic field strength of B335 was determined to be $30.2 \pm 17.7$ $\mu {\rm G}$. The critical mass of B335, evaluated using both magnetic and thermal/turbulent support against collapse, was determined to be $M_{\rm cr} = 3.37 \pm 0.94$ ${\rm M}_{\odot}$, which is identical to the observed core mass of $M_{\rm core}=3.67$ M$_{\odot}$. We thus concluded that B335 started its contraction from a condition near equilibrium. We found a linear relationship in the polarization versus extinction diagram, up to $A_V \sim 15$ mag toward the stars with the greatest obscuration, which verified that our observations and analysis provide an accurate depiction of the core. 
\end{abstract}

\vspace*{0.3 cm}

\clearpage

\section{Introduction}
Barnard 335 (B335) is one of the most well-defined isolated dense cores. The core harbors far-infrared (FIR) source IRAS 19347+0727, which is associated with a dense molecular gas envelope (e.g., Frerking et al. 1987; Menten et al. 1989; Saito et al. 1999; Kurono et al. 2013). The FIR source shows strong sub-millimeter (submm) emissions (Chandler et al. 1990) and is categorized as a Class 0 protostar based on its spectral energy distribution (Barsony 1994). The protostar is thought to be the driver of the bipolar outflow streaming in an east-west direction (e.g., Frerking \& Langer 1982; Hirano et al. 1988, 1992; Yen et al. 2010). Molecular line spectra display an inverse P-Cygni profile toward the central region of the core (e.g., Zhou et al. 1993; Choi et al. 1995; Evans et al. 2005, 2015), which is evidence of gas moving inwardly toward the core center, as expected for a protostar. The core is seen in optical images as a dark, mostly opaque globule that obscures background starlight and is elongated toward north-south direction (see Figure 3 of Galf\aa lk \& Olofsson 2007, see also Stutz et al. 2008 for the detection of flattened core). 
\par
Owing to its isolated geometry, simple shape, and rich stellar backdrop, the density structure of B335 was investigated with near-infrared (NIR) extinction measurements (Harvey et al. 2001). The Bonnor--Ebert sphere model (Ebert 1955; Bonnor 1956), which describes a pressure-confined, self-gravitating isothermal gas sphere in hydrostatic equilibrium, was well suited for describing the density structure of the core using a dimensionless radius of $\xi_{\rm max}=12.5 \pm 2.6$ (Harvey et al. 2001). This $\xi_{\rm max}$ value, which is a measure of the gas sphere's stability against gravitational collapse, clearly exceeds the stability criterion of 6.5 (Bonnor 1956), resulting in an unstable equilibrium solution. As shown by Kandori et al. (2005), the density profile of a collapsing gas sphere mimics a series of unstable Bonnor--Ebert solutions. The solution $\xi_{\rm max}=12.5$ is thus consistent with the existence of a protostar forming inside B335. Subsequent density structure observations at radio wavelengths suggested a density profile slope of $p \sim -1.5$ for the $r \propto \rho^p$ relationship in the inner region and $p \sim -2$ in the outer region (Harvey et al. 2003; Kurono et al. 2013). This is consistent with the shape of the density profile of an inside-out collapse solution (Shu 1977), as well as an unstable Bonnor--Ebert sphere with a rather flat inner region (plateau) and $p \sim -2$ profile in the outer region. Since these profiles cannot be discerned using observational data (Harvey et al. 2001), we treat the shape of density structure of B335 as a Bonnor--Ebert sphere with $\xi_{\rm max}=12.5$ in this study. 
\par
On the basis of the NIR extinction data and the distance obtained by Tomita et al. (1979), the mass of B335 inside its radius ($R=125''$, obtained by the Bonnor--Ebert fitting, Harvey et al. 2001) was determined to be $M_{\rm core}=14$ M${}_\odot$ ($d^2/250$ pc) (Harvey et al. 2001). Using the newly revised distance of $d=105 \pm 15$ pc (Olofsson \& Olofsson 2009) determined using the photometric distance of foreground and background stars, the mass and radius of B335 are $M_{\rm core}=3.67$ M${}_\odot$ and $R=13,100$au (0.0656 pc), respectively. 
\par
Zhou et al. (1990) reported the gas kinematic temperature $T_{\rm k}=13$ K and effective sound speed $C_{\rm s,eff}=0.23$ km s$^{-1}$ (i.e., turbulent velocity dispersion $\sigma_{\rm turb}=0.085$ km s$^{-1}$) for B335, through the microturbulent radiative transfer modeling of H$_2$CO 6 cm molecular line observations. Using the same parameter in theoretical modeling, gas infall spectra of various molecules were well explained (Zhou et al. 1993; Choi et al. 1995; Evans et al. 2005,2015). Note that NH$_3$ observations produced temperatures that were slightly higher ($T_{\rm k}=15$ K toward the center of the core, from Kurono et al. 2013) and lower ($T_{\rm k}=10-12$ K for the bulk of gas, from Menten et al. 1984). 
\par
The magnetic field is the last physical property of B335 that needs to be explored, and it has been the subject of several studies. Wolf et al. (2003) conducted submm dust emission polarimetry toward the core to obtain a magnetic field strength of $134^{+46}_{-39}$ $\mu$G and mean magnetic field direction of $\theta_{\rm mag}=3^{\circ}$. This magnetic field direction is almost perpendicular to the outflow axis. Bertrang et al. (2014) conducted NIR ($J_{\rm s}$ band) dust extinction polarimetry toward the core to obtain a magnetic field strength of $12-40$ $\mu$G and mean magnetic field direction of $\theta_{\rm mag}=115^{\circ} \pm 6^{\circ}$. Optical polarimetry of the B335 region showed a similar mean magnetic field direction of $\theta_{\rm mag}=111^{\circ} \pm 4^{\circ}$ (Vrba et al. 1986). Comparing these observations, mean magnetic field directions obtained at submm wavelengths are almost perpendicular to the direction obtained at optical or NIR wavelengths (note that recently Yen et al. 2019 reported that the core-scale magnetic field of B335 is in east-west direction, based on the submm polarimetry using JCMT/POL-2). Furthermore, the estimates of magnetic field strength vary, although the uncertainty in both observations are large. 
\par
There are several possible reasons for these differences. First, the field of view of the submm dust emission polarimetry was small, less than $1'$, so the submm results could be affected by the magnetic fields associated with the disk-like structure around the protostar. Furthermore, magnetic field strength was estimated using the data covering only the central region of the core, resulting in a large value, not an estimate of the mean magnetic field strength for the entire core. Second, the submm polarization map clearly contains a \lq \lq polarization hole''. The obtained submm polarization degree is small toward the center and gets larger toward the outer region. Therefore, it is anti-correlated with column density and does not reflect the overall magnetic field structure of the core. This may be due to the effect of optical depth and/or grain growth (Brauer et al. 2016). The same effect is seen in the subsequent dust emission polarimetry of B335 (Hull et al. 2014) and other dense cores (Matthews et al. 2009). 
Finally, though the NIR extinction polarimetry covered the entire core extent of $\sim$$4'$, the field of view was insufficient to map the outer ambient polarization distribution. Thus, the NIR data toward B335 are a superposition of the B335 polarization and the polarization arising from the ambient medium. This effect makes the observed polarization distribution of the core deviate from the true values. Note that optical polarimetry only traced ambient fields or the periphery of the core region. For these reasons, the magnetic field structure of B335 was not fully revealed based on the previous observations. 
\par
This study conducted wide-field background star linear polarimetry at NIR wavelengths for B335. The plane-of-sky magnetic field structure was revealed using several tens of stars in and around core radius, and the ambient field component was subtracted. The total magnetic field strength was estimated based on the Davis--Chandrasekhar--Fermi method (Davis 1951; Chandrasekhar \& Fermi 1953) and three-dimensional (3D) magnetic field modeling of the core. The magnetic field information was then used to determine the stability of B335 and describe the geometry among magnetic fields, outflow axis, and core elongation. The polarization versus extinction ($P$--$A$) relationship in B335 was constructed using corrections for (1) ambient polarization, (2) depolarization effect caused by a distorted magnetic field shape, and (3) line-of-sight magnetic inclination angle. The obtained linear $P$--$A$ relationship verifies that the polarizations reported here reflect the overall magnetic field structure in the core.

\section{Observations and Data Reduction}
We observed B335 using the $JHK_s$-simultaneous imaging camera SIRIUS (Nagayama et al. 2003) and its polarimetry mode SIRPOL (Kandori et al. 2006) on the Infrared Survey Facility (IRSF) 1.4-m telescope at the South African Astronomical Observatory (SAAO). IRSF/SIRPOL is one of the most useful instruments for NIR polarization surveys, providing deep- and wide- ($7.\hspace{-3pt}'7 \times 7.\hspace{-3pt}'7$ with a scale of 0$.\hspace{-3pt}''$45 ${\rm pixel}^{-1}$) field polarization images. The uncertainty due to sky variation during exposures is typically $0.3\%$ in polarization degree. The uncertainty of polarization angle due to the uncertainty in the determination of the polarization angle origin of the polarimeter is less than 3$^{\circ}$ (Kandori et al. 2006 and updates for Kusune et al. 2015). 
\par
We observed the polarized standard star RCrA\#88 ($P_H = 2.73\% \pm 0.07\%$, $\theta_H = 92^{\circ} \pm 1^{\circ}$, Whittet et al. 1992) on July 13, 2017 during the two months run of our observations. We obtained $P_H = 2.82\% \pm 0.09\%$ and $\theta_H = 91.9^{\circ} \pm 0.9^{\circ}$ which is consistent with the value in literature. 
\par
Observations were conducted on the nights of June 30 and July 2, 2017 using the linear polarimetry mode of SIRPOL. Ten-second exposures at four half-waveplate angles (in the sequence $0^{\circ}$, $45^{\circ}$, $22.5^{\circ}$, and $67.5^{\circ}$) were performed at 10 dithered positions (1 set). The total integration time was 2000 seconds (20 sets) per waveplate angle. The typical seeing during the observations was $\sim$$1.\hspace{-3pt}''35$ (3.0 pixels) in the $H$ band. \par
We reduced the observed data in the same manner as described in Kandori et al. (2007) using the Interactive Data Language (IDL) software (flat-field correction with twilight flat frames, median sky subtraction, and frame combination after registration). 
Note that the accuracy of calibration based on twilight flat is discussed in the Appendix of Kandori et al. (2020b). 
Software aperture polarimetry was carried out for a number of sources in the field of view. The point sources having a peak intensity greater than $10 \sigma $ above local sky background were detected on the Stokes $I$ image. 
The number of detected source is 2089, 2181, and 1385 in the $J$, $H$, and $K_s$ bands, respectively. The limiting magnitudes are 18.5, 18.0, and 17.2 mag in the $J$, $H$, and $K_s$ bands, respectively. 
The local background was subtracted using the mean of an annulus around the source on the original image. This process was carried out for each position angle image ($I_{0}$, $I_{45}$, $I_{22.5}$, and $I_{67.5}$). The aperture radius was same as the full width at half maximum of stars (3.0, 3.0, and 2.8 pixels for the $J$, $H$, and $K_s$ bands, respectively), and the sky annulus was set to 10 pixels with a 5-pixel width. This relatively small aperture radius was used to suppress flux contamination from neighboring stars in the crowded field. Another method to avoid stellar contaminations is the use of the psf-fitting photometry on each waveplate angle image. In this case, however, the goodness of psf-fitting for each star varies on each waveplate angle image, and this can be the source of systematic errors in measured polarization signals. Thus, we do not use the psf-fitting method for the polarization measurements of B335. Note that though we use relatively small radius for the aperture polarimetry, our important conclusions in this paper do not change even if we change the size of the aperture.
\par
The sources with a photometric error greater than 0.1 mag were ignored. The Stokes parameter for each star was obtained using $I = (I_{0} + I_{45} + I_{22.5} + I_{67.5})/2$, $Q = I_{0} - I_{45}$, and $U = I_{22.5} - I_{67.5}$. The polarization degree $P$ and polarization angle $\theta $ were then derived by $P = \sqrt{Q^2 + U^2}/I$ and $\theta = 0.5 {\rm atan}(U/Q)$. Since the polarization degree, $P$, is a positive quantity, the derived $P$ values tend to be overestimated, especially for low $S/N$ sources. We corrected the bias using $P_{\rm db} = \sqrt{P^2 - \delta P^2}$ (Wardle \& Kronberg 1974). The $H$ band observations provided polarizations for a total of 108 stars with $P/\delta P \ge 5$ in the field of view.
\par 
In this study, we used the results only for the $H$ band, where NIR extinction by dust is less severe than in the $J$ band and polarization efficiency is greater than that for the $K_s$ band. Note that the polarization vectors detected in the $J$ band and $K_s$ band are roughly coincident with those in the $H$ band. The correlation coefficients of $P/\delta P > 10$ sources for polarization angle are 0.62 ($J$ versus $H$) and 0.82 ($K_s$ versus $H$), and the coefficients for polarization degree are 0.79 ($J$ versus $H$) and 0.50 ($K_s$ versus $H$). 

\section{Results and Discussion}
\subsection{Distortion of Magnetic Fields}
Figure 1 shows the observed polarization vector map of B335 in the $H$ band. B335 is visible near the center of the image as a dark region that obscures the stars behind it. The white circle marks the core boundary ($R=125''$, Harvey et al. 2001). Inside the core radius, the bent structure of the polarization vectors can be seen, particularly in the northern part of the core. Outside of the core radius (i.e., the off-core region), polarization vectors generally flow from north-west to south-east. These polarization components can be regarded as off-core polarization, located along the same line of sight but unrelated to the B335 core. 
\par
Considering the direction of galactic longitude (${\rm PA} \sim 30^{\circ}$ in the equatorial coordinates at the position of B335), the off-core vectors deviate from the polarization direction of the galactic plane. As reported by Frerking et al. (1987), the B335 core is accompanied by diffuse envelope extending more than 20$'$ in east--west direction (see, Figure 1 of Frerking et al. 1987). The position angle of off-core vectors is roughly perpendicular to the elongation axis of the diffuse envelope, and may be associated with this structure.
\par
The off-core vectors have $P_{H,{\rm off}} = 1.43 \pm 0.49$\% and $\theta_{H,{\rm off}} = 141.9^{\circ} \pm 11.8^{\circ}$ (Figures 2 and 3). The off-core polarizations are relatively well ordered (Figure 3) and have similar polarization degree (Figure 2). Following the same method described in our previous papers (Kandori et al. 2017a, hereafter Paper I), we fitted the off-core vectors onto the sky plane using the Stokes parameters $Q/I$ and $U/I$. The distributions of $Q/I$ and $U/I$ values were modeled as $f(x,y)=A + Bx + Cy$, where $x$ and $y$ are the pixel coordinates, and $A$, $B$, and $C$ are the parameters to be fitted. The estimated off-core vectors are shown in Figure 4. The regression vectors of off-core components were subtracted from the original vectors to isolate the polarization vectors associated with B335 (Figure 5). Comparing Figures 1 and 5, polarization vectors toward B335 are slightly rotated by the effect of ambient subtraction. After subtraction, polarization degrees of off-core vectors were successfully suppressed toward 0\% (Figure 2), and approximately random distribution of polarization angles was realized (Figure 3). 
\par
Some on-core polarization vectors increase in polarization degree after the subtraction of off-core vectors. This occurs for the on-core vectors with angles perpendicular to the off-core vectors, and is not an artificial effect. The increase of polarization degree for some on-core vectors is the result of the suppression of the depolarization effect by off-core vectors.
\par
In Figure 5, the north part of off-core polarizations is not fully subtracted out and seems to be associated with the polarization pattern of the core region. As described above, these vectors may be associated with the diffuse envelope structure surrounding the B335 core.
\par
In Figure 5, polarization vectors toward B335 are clearly oriented east-west, and a structure reflecting the bending of magnetic fields can be seen in the northern and (possibly) southern part of the core. The number of polarization vectors within the core radius of $R \le 125''$ and with $P_H \ge 1$\% is 24. This number is small, but sufficient to draft the magnetic field lines. 
%

\subsection{Parabolic Model}
The most probable configuration of the magnetic field lines, estimated using a parabolic function and its rotation, is shown in Figure 6 (solid white lines). The field of view matches the diameter of B335 ($250''$), and the 24 polarization vectors having $P_H \ge 1$\% are shown in the figure. 
The threshold of $P_H \ge 1$\% was determined to avoid the systematic error caused by the subtraction of off-core polarizations. Note that most of the residual polarization vectors in off-core region after the subtraction analysis are less than 1\%, as shown in Figure 2. The coordinate origin of the parabolic function is fixed to the central position of the core/protostar system measured on the Atacama Large Millimeter Array (ALMA) dust emission map (R.A.=19$^{\rm h}$37$^{\rm m}$00$.\hspace{-3pt}^{\rm s}$89, Decl.=+7$^{\circ}$34$'$10$.\hspace{-3pt}''0$, J2000, Evans et al., 2015).
\par
The best-fit parameters are $\theta_{\rm mag}=90^{\circ} \pm 7^{\circ}$ and $C = 1.43 (\pm 0.49) \times 10^{-5}$ ${\rm pixel}^{-2}$ ($= 7.06 \times 10^{-5}$ ${\rm arcsec}^{-2}$) for the parabolic function $y = g + gC{x^2}$, where $g$ specifies magnetic field lines, $\theta_{\rm mag}$ denotes the position angle of the magnetic field direction (from north through east), and $C$ determines the degree of curvature in the parabolic function. 
In the fitting procedure, the observational error for each star was taken into account in the calculations of $\chi^2 = (\sum_{i=1}^n (\theta_{\rm obs,{\it i}} - \theta_{\rm model}(x_i,y_i))^2 / \delta \theta_i^2$, where $n$ is the number of stars, $x$ and $y$ are the star coordinates, $\theta_{\rm obs}$ and $\theta_{\rm model}$ denote the polarization angle from observations and the model, respectively, and $\delta \theta_i$ is the observational error). 
The parabolic fitting seems reasonable, since the standard deviation of residual angles, $\theta_{\rm res} = \theta_{\rm obs} - \theta_{\rm fit}$, is smaller for the parabolic function ($\delta \theta_{\rm res} = 21.54^{\circ} \pm 0.99^{\circ}$, Figure 7) than for the uniform field case ($\delta \theta_{\rm uni} = 24.94^{\circ} \pm 1.00^{\circ}$). 
When $\delta \theta_{\rm res}$ and $\delta \theta_{\rm uni}$ are compared, the difference is statistically significant at the 3-sigma level, and the existence of the distorted field is most likely real. 
\par
B335 is the first protostellar core to be associated with axisymmetrically distorted hourglass-shaped magnetic fields, and is the third dense core to be associated with such fields following the results of FeSt 1-457 (Paper I) and B68 (Kandori et al., 2019), which lack protostars. The obtained magnetic curvature value $C= 7.06 \times 10^{-5}$ ${\rm arcsec}^{-2}$ $=6.4 \times 10^{-9}$ ${\rm AU}^{-2}$ is similar to that obtained for FeSt 1-457 ($C = 5.14 \times 10^{-5}$ ${\rm arcsec}^{-2}$ $=3.0 \times 10^{-9}$ ${\rm AU}^{-2}$, Paper I) and B68 ($C = 1.09 \times 10^{-4}$ ${\rm arcsec}^{-2}$ $=7.0 \times 10^{-9}$ ${\rm AU}^{-2}$, Kandori et al., 2019), suggesting the existence of similar mechanisms that create hourglass-shaped field distortions in dense cores/globules. 
The magnetic curvature of dense cores can be produced with the drag of magnetic fields by materials concentrating toward center, and the degree of curvature can be determined by the moving distance of mass which drags magnetic fields. The similar curvature values among the dense cores indicate that the spatial scale of core formation is similar in these cores. Core formation scale, i.e., initial contraction radius $R_0$, may be similar in dense cores/globules. Further observational studies to measure magnetic curvature of dense cores are needed. 

\par
The intrinsic dispersion, $\delta \theta_{\rm int} = (\delta \theta_{\rm res}^2 - \delta \theta_{\rm err}^2)^{1/2}$, estimated using the parabolic fitting, is $21.34^{\circ} \pm 1.00^{\circ}$ (0.3725 radian), where $\delta \theta_{\rm err}$ is the standard deviation of the observational error in polarization measurements. Note that the choice of a function form better than the function used here (i.e., parabolic form) can reduce the dispersion at residual angles (e.g., a mathematical model for hourglass fields by Ewertowski \& Basu 2013, Myers et al. 2018).
\par
%
%
If the magnetic field is assumed to be frozen in the medium, the intrinsic dispersion of the magnetic field direction, $\delta \theta_{\rm int}$, can be attributed to an Alfv\'{e}n wave perturbed by turbulence. The strength of the plane-of-sky magnetic field (${B}_{\rm pos}$) can be estimated from the relation ${B}_{\rm pos} = {C}_{\rm corr} (4 \pi \rho)^{1/2} \sigma_{\rm turb} / \delta \theta_{\rm int}$, where $\rho$ and $\sigma_{\rm turb}$ are the mean density of the core and turbulent velocity dispersion, respectively (Davis 1951; Chandrasekhar \& Fermi, 1953) and ${C}_{\rm corr}$ is a correction factor suggested by theoretical studies. In the original formulation, ${C}_{\rm corr} = 1$ (Davis 1951; Chandrasekhar \& Fermi, 1953), whereas in this study, a value of ${C}_{\rm corr} = 0.5$ (Ostriker et al., 2001, see also, Padoan et al. 2001; Heitsch et al. 2001; Heitsch 2005; Matsumoto et al. 2006) was adopted. 
Using the data for mean density $(\rho = 2.30 (\pm 1.74) \times 10^{-19}$ g cm$^{-3}$) calculated from Harvey et al. (2001), turbulent velocity dispersion ($\sigma_{\rm turb} = 0.085$ km s${}^{-1}$) from Zhou et al. (1990), and $\delta \theta_{\rm int}$ derived here, a relatively weak magnetic field was obtained as a lower limit of total field strength ($|B|$): $B_{\rm pos} = 19.4 \pm 11.4$ $\mu {\rm G}$. 
\par
We compared the obtained plane-of-sky magnetic field strength with previous studies ($134^{+46}_{-39}$ $\mu$G, Wolf et al. 2003; $12 - 40$ $\mu$G, Bertrang et al. 2014). Note that these previous studies used no theoretical correction factor in the Davis--Chandrasekhar--Fermi calculation. If we apply $C_{\rm corr} = 0.5$ to their data, the values of the magnetic field strength are reduced to $67^{+23}_{-20}$ $\mu$G (Wolf et al. 2003) and $6 - 20$ $\mu$G (Bertrang et al. 2014), respectively. The value by Bertrang et al. (2014) is roughly coincident with our result, and the value by Wolf et al. (2003) is larger than our result. Note that the submm polarimetry by Wolf et al. (2003) covered only the central region of the core, and the obtained magnetic field strength is not the mean value for the entire core. 

\subsection{3D Magnetic Field}
For the 3D magnetic field modeling, we followed the procedure described in our previous papers (Kandori et al. 2017b, hereafter Paper II, see also Kandori et al. 2020a, hereafter Paper VI). The 3D version of the simple parabolic function employed in Paper I, $z(r, \varphi, g) = g + gC{r}^{2}$ in the cylindrical coordinate $(r, z, \varphi)$, was used for modeling the core magnetic fields, where $g$ specifies the magnetic field line, $C$ is the curvature of lines, and $\varphi$ is the azimuth angle (measured in the plane perpendicular to $r$). In this function, the shape of magnetic field lines is axially symmetric around the $r$ axis. Thus, the function $z(r, \varphi, g)$ has no dependence on the parameter $\varphi$. 
\par
The 3D model was virtually observed after rotating in the line-of-sight ($\gamma_{\rm mag}$) and the plane-of-sky ($\theta_{\rm mag}$) directions. The analysis was performed in the same manner described in Section 3.1 of Paper VI (see also Sections 2 and 3.1 of Paper II). The resulting polarization vector maps for the 3D parabolic model for various viewing angles ($90^{\circ} - \gamma_{\rm mag}$) are shown in Figure S1 of Paper VI. The density distribution of the model core was calculated based on the Bonnor--Ebert sphere model (Ebert 1955; Bonnor 1956) with a solution parameter of $\xi_{\rm max} = 12.5$ (Harvey et al. 2001). Since we obtained a linear polarization--extinction relationship for B335 (see Section 3.5), it is reasonable to assume that the polarization arisen in a cell in the model core is proportional to the density in the cell. The model map can be compared with observations for each parameter set (see Figure S1 of Paper VI). Since the polarization distributions in the model core differ from one another depending on the viewing angle toward the line of sight, $\chi^2$ fitting of these distributions with the observational data can be used to restrict the line-of-sight magnetic inclination angle and 3D magnetic curvature. Note that the resolution of the final model 3D map is enough for comparison with observations. The nearest neighbor distance of stars in observations is 44 pixels on the average, whereas the sampling grid width of the model is 9 pixels ($\approx 4''$). 
\par
Figure 8 summarizes the distribution of $\chi^2_\theta = ( \sum_{i=1}^n (\theta_{\rm obs,{\it i}} - \theta_{\rm model}(x_i,y_i))^2 / \delta \theta_i^2$, where $n$ is the number of stars, $x$ and $y$ show the coordinates of stars, $\theta_{\rm obs}$ and $\theta_{\rm model}$ denote the polarization angle from observations and the model, and $\delta \theta_i$ is the observational error) calculated by using the model and observed polarization angles. The optimal magnetic curvature parameter, $C$, was determined at each inclination angle $\gamma_{\rm mag}$ to obtain $\chi^2_\theta$. In Figure 8, the minimization point is $\gamma_{\rm mag} = 45^{\circ}$. It is clear that the high-$\gamma_{\rm mag}$ region (especially $\gamma_{\rm mag} \simgt 80^{\circ}$) is unlikely. Note that the large $\gamma_{\rm mag}$ model approaches the pole-on magnetic field geometry and shows a radial polarization pattern that does not match observations. 
\par
To check consistency of the $\chi^2$ analysis based on polarization angle, we made $\chi^2$ calculations using both polarization angles and degrees. We determined the best model parameters for each $\gamma_{\rm mag}$ by minimizing the difference in polarization angles, and we calculated $\chi^2_P = (\sum_{i=1}^n (P_{\rm obs,{\it i}} - P_{\rm model}(x_i,y_i))^2 / \delta P_i^2$, where $P_{\rm obs}$ and $P_{\rm model}$ show the polarization degree from observations and the model, and $\delta P_i$ is the observational error) in polarization degrees (Figure 9). In the procedure, the relationship between the model core column density and polarization degree was scaled to be consistent with observations. In Figure 9, the minimization point is $\gamma_{\rm mag} = 55^{\circ}$. It is clear that the low-$\gamma_{\rm mag}$ region (especially $\gamma_{\rm mag} \simlt 30^{\circ}$) is unlikely. Note that the small $\gamma_{\rm mag}$ model approaches the edge-on magnetic field geometry and shows a hourglass-like field structure with weak depolarization pattern. 
\par
The distributions of $\chi^2_{\theta}$ and $\chi^2_{P}$ show relatively large scatter, especially for small or large $\gamma_{\rm mag}$. This can be due to the complicated patterns of hourglass-like fields projected on to the skyplane, and the relatively small number of data points ($N=24$) of the B335's background stars. 
\par
The 1-sigma error estimated at the minimum $\chi^2_{\theta}$ and $\chi^2_{P}$ points are $12.5^{\circ}$ and $4.7^{\circ}$, respectively. These values are similar to the difference between $\chi^2$ minimization points based on $\chi^2_{\theta}$ and $\chi^2_{P}$ ($10^{\circ}$). We conclude that the most likely value for $\gamma_{\rm mag}$ is $50^{\circ}$ with uncertainty of $10^{\circ}$. The magnetic curvature obtained at $\gamma_{\rm mag} = 50^{\circ}$ is $C = 1.73 \times 10^{-4}$ ${\rm arcsec}^{-2}$. Note that the $\chi^2$ values in Figures 8 and 9 are relatively large. This seems to come from the existence of polarization angle scattering mainly caused by Alfv\'{e}n waves, which cannot be included in the observational error term in the calculation of $\chi^2$. 
\par
Figure 10 shows the best-fit 3D parabolic model with the observed polarization vectors. The direction of polarization vectors in the model generally agrees with observations at the same level compared with the results of 2D fitting, although there is some deviations in between model and observations particularly in the north part of the core. The standard deviation of the angular difference in plane-of-sky polarization angles between the 3D model and observations is $15.12^{\circ}$, which is less than the fitting result in 2D, as well as in the uniform field case of $\delta \theta_{\rm uni} = 24.94^{\circ}$. 
\par
Figure 11 is the same observational data as Figure 10, but with the background image processed using the line integral convolution (LIC) technique (Cabral \& Leedom, 1993). We used publicly available IDL code developed by Diego Falceta-Gon\c{c}alves. The direction of the LIC \lq \lq texture'' is parallel to the direction of magnetic fields, and the background image is based on the polarization degree of model core. 

\subsection{Magnetic Properties of the Core}
Using the magnetic inclination angle $\theta_{\rm inc}$ of $50^{\circ} \pm 10^{\circ}$ obtained as described in the previous section, we determined that the mean magnetic field strength of B335 is $B_{\rm pos}/\cos (\gamma_{\rm mag})=19.4/\cos(50^{\circ})=30.2 \pm 17.7$ $\mu{\rm G}$. 
The ability of the magnetic field to support the core against gravity was investigated using the parameter ${\lambda} = ({M}/{\Phi})_{\rm obs} / ({M}/{\Phi})_{\rm critical}$, which represents the ratio of the observed mass-to-magnetic flux ratio to a critical value, $(2\pi {\rm G}^{1/2})^{-1}$, suggested by theory (Mestel \& Spitzer 1956; Nakano \& Nakamura 1978). We found that $\lambda = 3.24 \pm 1.52$ (magnetically supercritical) and that the magnetic critical mass of the core is $1.13 \pm 0.69$ ${\rm M}_{\odot}$, which is lower than the observed core mass of $M_{\rm core}=3.67$ ${\rm M}_{\odot}$. 
\par
The critical mass of B335, evaluated using both magnetic and thermal/turbulent support against collapse, $M_{\rm cr} \simeq M_{\rm mag}+M_{\rm BE}$ (Mouschovias \& Spitzer 1976; Tomisaka, Ikeuchi, \& Nakamura 1988; McKee 1989), is $1.13+2.24=3.37 \pm 0.94$ ${\rm M}_{\odot}$, where $2.24 \pm 0.64$ ${\rm M}_{\odot}$ is the Bonnor--Ebert mass calculated using the kinematic temperature of the core of 13 K, turbulent velocity dispersion of $0.085$ km ${\rm s}^{-1}$ (Zhou et al. 1990), and external pressure $P_{\rm ext}$ of $1.35(\pm 0.77) \times 10^5$ K cm$^{-3}$ calculated from Harvey et al. (2001). Though the obtained $M_{\rm cr}$ is slightly less than the core mass $M_{\rm core}$, we concluded that the stability of B335 is near the critical state, with $M_{\rm cr} \sim M_{\rm core}$. 
\par
Thus, the protostellar core B335 appears to have started its contraction from a condition near equilibrium between gravity and magnetic and thermal/turbulent support. We speculate that, in general, the (spontaneous) low-mass star formation in globules is initiated from the state close to the critical condition. This is consistent with the observations of subsonic infall motion in dense cores (Lee et al. 1999, 2001). If the formation of stars were to start from a condition far out of equilibrium, the dense core would collapse suddenly with strong acceleration, and supersonic infall motion would be widely observed toward dense cores (see, e.g., Aikawa et al. 2005). The infalling gas motion in B335 was successfully modeled (e.g., Zhou et al. 1993; Choi et al. 1995; Evans et al. 2005, 2015) using the spherical inside-out collapse model (Shu 1977), in which the collapse starts from the end of quasi-static evolution. This is consistent with our conclusion that B335 is in a nearly equilibrium state. 
\par
We evaluated the stability parameter $\lambda_{\rm cr} \equiv M_{\rm core}/M_{\rm cr} = M_{\rm core}/(M_{\rm mag}+M_{\rm BE})$, and found that $\lambda_{\rm cr}$ is 1.09 for B335, which can be compared to $\lambda_{\rm cr}$ of 0.94 for FeSt 1-457 (Paper I,II,VI) and 0.91 for B68 (Kandori et al. 2019). Two starless cores and a protostellar core all show $\lambda_{\rm cr}$ values close to unity, again supporting our conclusion that the initial condition of star formation in isolated globules may be in the condition very close to the critical state. If this is true, a slight decrease in turbulence (Nakano 1998) and/or ambipolar diffusion can initiate the onset of star formation. However, the formation mechanism of nearly critical dense cores/globules remains unknown and is an open problem. 
\par
The relative importance of magnetic fields in supporting the core against gravity was investigated using the ratio of the thermal and turbulent energy to the magnetic energy, ${\beta} \equiv 3{C}_{\rm s}^{2}/{V}_{\rm A}^{2}$ and ${\beta}_{\rm turb} \equiv {\sigma}_{\rm turb,3D}^{2}/{V}_{\rm A}^{2} = 3{\sigma}_{\rm turb,1D}^{2}/{V}_{\rm A}^{2}$, where ${C}_{\rm s}$, ${\sigma}_{\rm turb}$, and ${V}_{\rm A}$ denote the isothermal sound speed at 13 K, turbulent velocity dispersion, and Alfv\'{e}n velocity, respectively. 
These ratios were found to be ${\beta} \approx 4.39$ and ${\beta}_{\rm turb} \approx 0.69$, respectively. Thus, B335 is dominated by thermal support with a smaller contribution from static magnetic fields. Turbulence seems dissipated and makes a contribution comparable with or smaller than that of magnetic fields. 
These values were previously derived for FeSt 1-457 $({\beta} \approx 1.27$ and ${\beta}_{\rm turb} \approx 0.12$ with the application of the core's inclination angle) (Paper I; Paper II) and B68 (${\beta} \approx 3.15$ and ${\beta}_{\rm turb} \approx 0.70$, Kandori et al. 2019). 
Comparison of the results for all three dense cores indicates that they are mainly thermally supported with comparably smaller contributions from magnetic fields. Though the support from turbulence seems to be minor, this does not necessarily mean that turbulence is unimportant. Considering the nearly critical state of dense cores, even slight further dissipation of turbulence can initiate core contraction (e.g., Nakano 1998). 
\par
Figure 12 shows the relationship among the core's elongation axis ($\theta_{\rm elon} \sim 0^{\circ}$), outflow ($\theta_{\rm out} \sim 90^{\circ}$), and magnetic field ($\theta_{\rm mag}=90^{\circ}$) superimposed on the map obtained by submm dust emission polarimetry (Wolf et al. 2003). Like FeSt 1-457 and B68, the orientation of the elongation of the optical obscuration of B335, as well as in the H$^{13}$CO$^{+}$ ($J=1-0$) line map (Kurono et al. 2013), is clearly perpendicular to the magnetic axis. This geometrical relationship is consistent with the picture of mass accretion along magnetic field lines suggested by theories (e.g., Galli \& Shu 1993a,b). This geometry is also consistent with the magneto-hydrostatic configuration (e.g., Tomisaka, Ikeuchi, \& Nakamura 1988) as the starting condition for core contraction. The outflow axis is toward east-west (Hirano et al. 1988) on the plane of sky and is nearly perpendicular to the line of sight, with an inclination angle of 87$^{\circ}$ (Stutz et al. 2008). Though the outflow axis and magnetic axis are aligned perfectly on the plane of sky, the magnetic inclination angle of $\gamma_{\rm mag}=50^{\circ} \pm 10^{\circ}$ means that gap probably exists between the two axes in the line of sight. 

\subsection{Polarization--Extinction Relationship}
As described in the previous paper (Kandori et al. 2018a, hereafter Paper III, see also Paper VI), the relationship between dust dichroic polarization ($P$) and extinction ($A$) in molecular clouds and their cores is important for (1) interpreting the obtained angle of interstellar polarizations, which is closely related to the direction of magnetic fields pervading the observed region, and (2) investigating dust alignment mechanisms in the dense environment. The former point is essential to ensure that our polarization observations can trace the dust alignment and magnetic fields deep inside B335. 
The latter point is important for the comparison of observations with dust alignment theories. 
\par
The observed polarizations toward B335 are a superposition of the polarization from the core and the ambient medium surrounding the core. The polarizations arising from the ambient off-core medium must be subtracted from the observed polarizations in order to extract the polarization associated with the core (Section 3.1). As described in Section 3.3, distorted magnetic fields surrounding the core can produce depolarization, particularly at the equatorial plane of the core. Furthermore, line-of-sight inclination angle in magnetic axis weakens the observed plane-of-sky polarizations. These effects must be corrected in order to determine the true relationship between polarization and extinction in dense cores. 
\par
Figure 13(a) shows the observed $P_H$ versus $H-K_{s}$ relationship with no correction. The stars with $R \le 125''$ and $P_H / \delta P_H \ge 6$ were used. We did not employed the threshold of $P_H \ge 1\%$ as used in the parabolic fit analysis in Section 3.2. Though several stars with $P_H \ge 1\%$ were included in the plot, this does not change the important conclusions. Note that the strongest polarization vector in Figure 1 was dropped in Figure 13(a) due to the large uncertainty in $P_H$. 
%
In the figure, the polarization generally increases with increasing extinction up to $H-K_{s} \sim 1.2$ mag. The slope of the relationship is $3.30 \pm 0.14$ $\%$ ${\rm mag}^{-1}$. The observed polarization is the superposition of the polarization from the core and ambient medium. Stars located outside of the core radius, $R>125''$, were used to estimate the off-core polarization vectors. After subtraction, a relatively linear $P_H$ versus $H-K_{s}$ relationship was obtained, as shown in Figure 13(b). The slope of the relationship is $4.67 \pm 0.17$ $\%$ ${\rm mag}^{-1}$, which is close to the average interstellar polarization slope (Jones 1989). Note that, since the ambient vectors are roughly north-south and the B335 vectors are roughly east-west, the slope of the $P$--$A$ relationship after subtraction is greater than the original value. 
\par
B335 is associated with inclined distorted magnetic fields, as discussed in Sections 3.1-3.3, which provides depolarization effects inside the core. Based on the known 3D magnetic field structure, a depolarization correction factor was estimated, as shown in Figure 14. This figure was created simply by dividing the polarization degree map of the distorted field model with edge-on geometry ($\gamma_{\rm mag} = 0^{\circ}$) by the inclined distorted field model ($\gamma_{\rm mag} = 50^{\circ}$). The correction factor map can thus simultaneously correct for both the depolarization and line-of-sight inclination angles. In Figure 14, the factors distributed around the equatorial plane (north-south direction) are less than unity (indicating depolarization). This is due to the crossing of the polarization vectors at the front and back sides of the core along the line-of-sight (see the explanatory illustration in Figure 7 of Kataoka et al. 2012). 
\par
Figure 13(c) shows the depolarization and inclination corrected $P$--$A$ relationship obtained by dividing the Figure 13(b) relationship by the correction factor map (Figure 14). The slope of the relationship is $12.73 \pm 0.43$ $\%$ ${\rm mag}^{-1}$. The final corrected slope is larger than those obtained for FeSt 1-457 ($6.60\pm0.41$ $\%$ ${\rm mag}^{-1}$, Paper III and Paper IV) and B68 ($4.57 \pm 0.11$ $\%$ ${\rm mag}^{-1}$, Kandori et al. 2019), and it is comparable to the statistically estimated upper limit of interstellar polarization efficiency (Jones 1989, $P_H / E_{H-K_s} \sim 14$). 
\par
The correlation coefficients obtained in these analysis are 0.80 (ambient polarization subtraction, Figure 13(b)) and 0.87 (ambient polarization, depolarization, and inclination correction, Figure 13(c)). These are comparable to the value of 0.85 for the original relationship. For B335, our analysis did not greatly improve the tightness of the $P$--$A$ correlation, but it changed the slope value. 
\par
Comparison of Figures 13(a) and 13(c) shows that the corrections produced a dramatic change, particularly in slope. The final result is a relatively linear relationship between polarization and extinction in the range up to $A_V \sim 15$ mag. This linear relationship verifies that our NIR polarimetric observations trace the overall polarization (magnetic field) structure of B335. 
\par
Figure 15 shows the $P_H / A_V$ versus $A_V$ diagram. Reflecting the relatively linear slope in Figures 13(c), distributions of data points seems flat especially for $A_V \simgt 5$ mag. To reject too high $P_H / A_V$ data for very low $A_V$, the data points with $P_H / A_V > 4$ \% mag$^{-1}$ were not plotted. The dotted line shows the fitting of the data using the power law $P_H / A_V \propto A_{V}^{\alpha}$, resulted in the $\alpha$ index of $-0.22 \pm 0.22$. 
The relatively shallow $\alpha$ index indicates that the polarization (and magnetic field) structure toward B335 is traced in our observations. The dotted-dashed line shows an observational upper limit by Jones (1989). The relation was calculated based on the equation $P_{K,{\rm max}} = \tanh{\tau_{\rm p}}$, where $\tau_{\rm p} = (1-\eta)\tau_{K}/(1+\eta)$, and the parameter $\eta$ is set to 0.875 (Jones 1989). $\tau_{K}$ denotes the optical depth in the $K$ band. The polarization efficiency of B335 is quite high, as high as the interstellar upper limit value. 

\section{Summary and Conclusion}
This study revealed the detailed magnetic field structure of the dense protostellar core B335 based on NIR polarimetric observations of background stars to measure dichroically polarized light produced by aligned dust grains in the core. After subtracting ambient polarization components, the magnetic fields pervading B335 were mapped using 24 stars, and axisymmetrically distorted hourglass-shaped magnetic fields were identified for the first time in a protostellar core. On the basis of simple 2D and 3D magnetic field modeling, magnetic inclination angles in the plane-of-sky and line-of-sight directions were determined as $90^{\circ} \pm 7^{\circ}$ and $50^{\circ} \pm 10^{\circ}$, respectively. We used the obtained magnetic inclination angle and the Davis--Chandrasekhar--Fermi method to calculate the total magnetic field strength of B335 as $B_{\rm pos}/ \cos(\gamma_{\rm mag}) = 19.4/ \cos(50^{\circ}) = 30.2 \pm 17.7$ $\mu$G. The magnetic critical mass of the core, $M_{\rm mag}=1.13 \pm 0.69$ M$_{\odot}$, is less than the observed core mass, $M_{\rm core}=3.67$ M$_{\odot}$, suggesting a magnetically supercritical state with a ratio of observed mass-to-flux ratio to the critical value $\lambda = 3.24 \pm 1.52$. The critical mass of B335, evaluated using both magnetic and thermal/turbulent support, $M_{\rm cr} \simeq M_{\rm mag}+M_{\rm BE}$, is $1.13+2.24=3.37 \pm 0.94$ ${\rm M}_{\odot}$, where $2.24 \pm 0.64$ ${\rm M}_{\odot}$ is the Bonnor--Ebert mass calculated using the effective sound speed and external pressure. Though the obtained $M_{\rm cr}$ is slightly less than the core mass $M_{\rm core}$, we concluded that the stability of B335 is in a condition near the critical state, with $M_{\rm cr} \sim M_{\rm core}$. Thus, B335 is considered to have started its contraction from the condition near the equilibrium. Further, we speculate that the (spontaneous) low mass star formation in globules is generally initiated in the state close to the critical condition. The direction of plane-of-sky magnetic fields of B335 ($\sim 90^{\circ}$) is perpendicular to the core's elongation axis ($\sim 0^{\circ}$), and is parallel to the outflow axis ($\sim 90^{\circ}$). This geometrical relationship is consistent with the picture of mass accretion along magnetic field lines suggested by theories of isolated star formation. We found a linear relationship in the polarization versus extinction diagram, up to $A_V \sim 15$ mag toward the stars with greatest obscuration. The linear relationship indicates that the observed polarizations reflect the actual overall magnetic field structure of the core. However, this linear nature also raises questions about the alignment of dust grains in the cold, dense, and low radiation environment of B335. Further theoretical and observational studies are needed to explain the dust alignment.

\bigskip

We are grateful to the staff of SAAO for their kind help during the observations. We wish to thank Tetsuo Nishino, Chie Nagashima, and Noboru Ebizuka for their support in the development of SIRPOL, its calibration, and its stable operation with the IRSF telescope. The IRSF/SIRPOL project was initiated and supported by Nagoya University, the National Astronomical Observatory of Japan, and the University of Tokyo in collaboration with South African Astronomical Observatory under the financial support of Grants-in-Aid for Scientific Research on Priority Area (A) Nos. 10147207 and 10147214, and Grants-in-Aid Nos. 13573001 and 16340061 of the Ministry of Education, Culture, Sports, Science, and Technology of Japan. RK, MT, NK, KT (Kohji Tomisaka), and MS also acknowledge support by additional Grants-in-Aid Nos. 16077101, 16077204, 16340061, 21740147, 26800111, 16K13791, 15K05032, 16K05303, 19K03922.

\clearpage 
\begin{figure}[t]
\begin{center}
 \includegraphics[width=6.5 in]{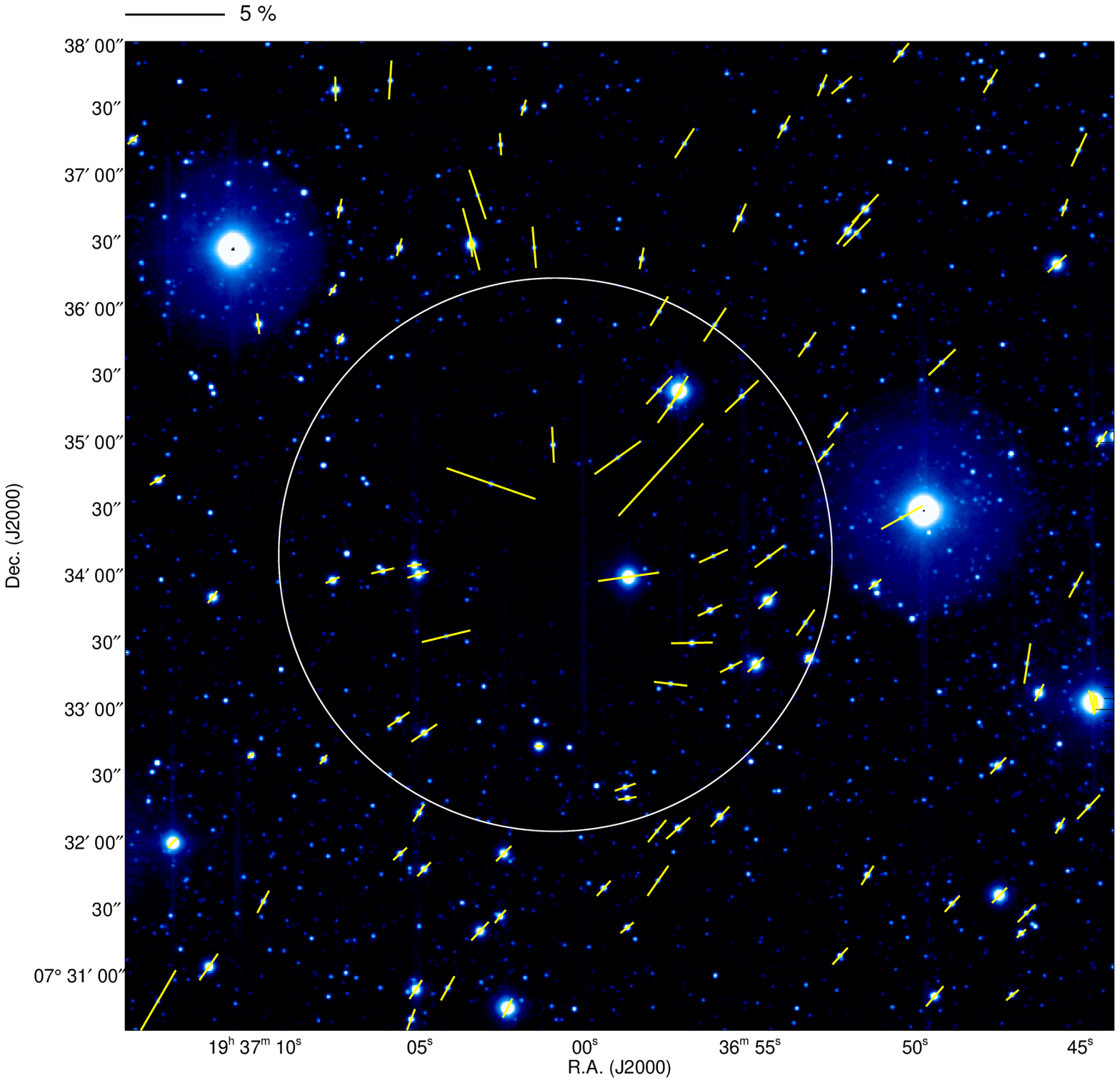}
\end{center}
 \caption{Polarization vectors of point sources superimposed on the $H$ band intensity image for B335. The white circle marks the core boundary (radius of 125$''$, Harvey et al. 2001). The scale of the 5\% polarization degree is shown above the image.}
   \label{fig1}
\end{figure}

\clearpage 
\begin{figure}[t]
\begin{center}
 \includegraphics[width=6.5 in]{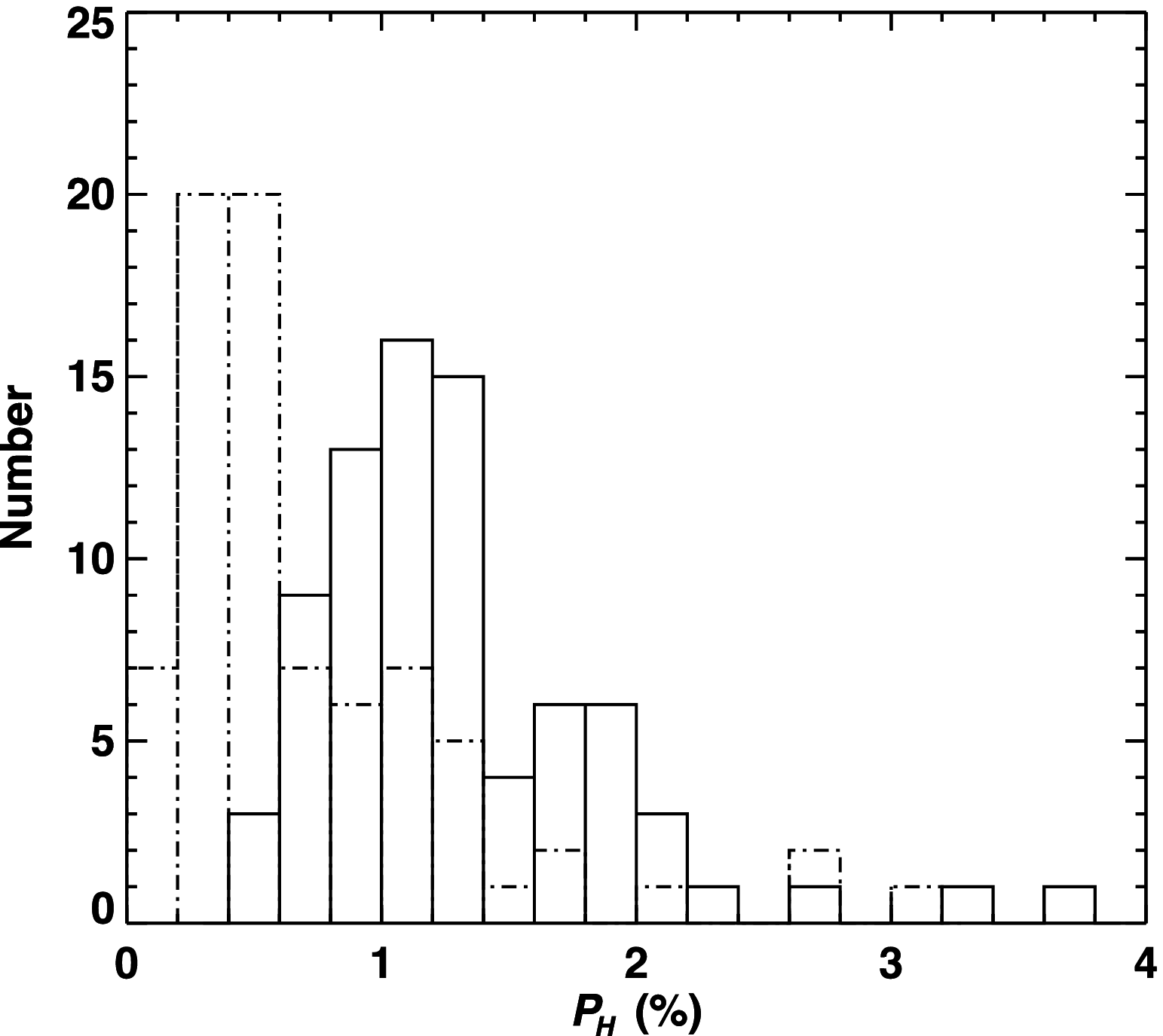}
\end{center}
 \caption{Histogram of $P_H$ for the stars in the off-core region before (solid line) and after (dotted-dashed line) subtraction of the off-core component.}
   \label{fig1}
\end{figure}

\clearpage 
\begin{figure}[t]
\begin{center}
 \includegraphics[width=6.5 in]{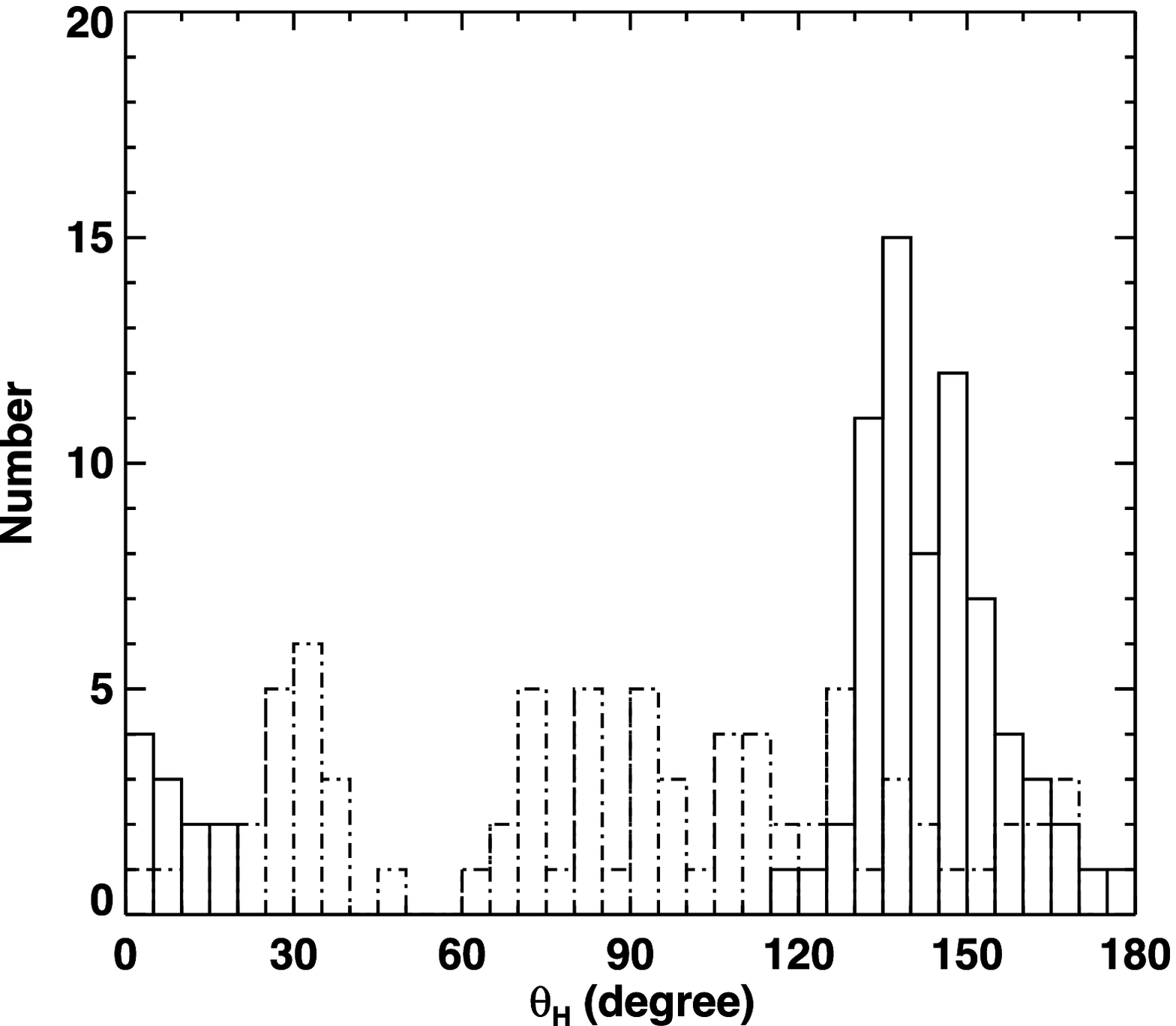}
\end{center}
 \caption{Histogram of $\theta_H$ for the stars in the off-core region before (solid line) and after (dotted-dashed line) subtraction of the off-core component.}
   \label{fig1}
\end{figure}

\clearpage 
\begin{figure}[t]
\begin{center}
 \includegraphics[width=6.5 in]{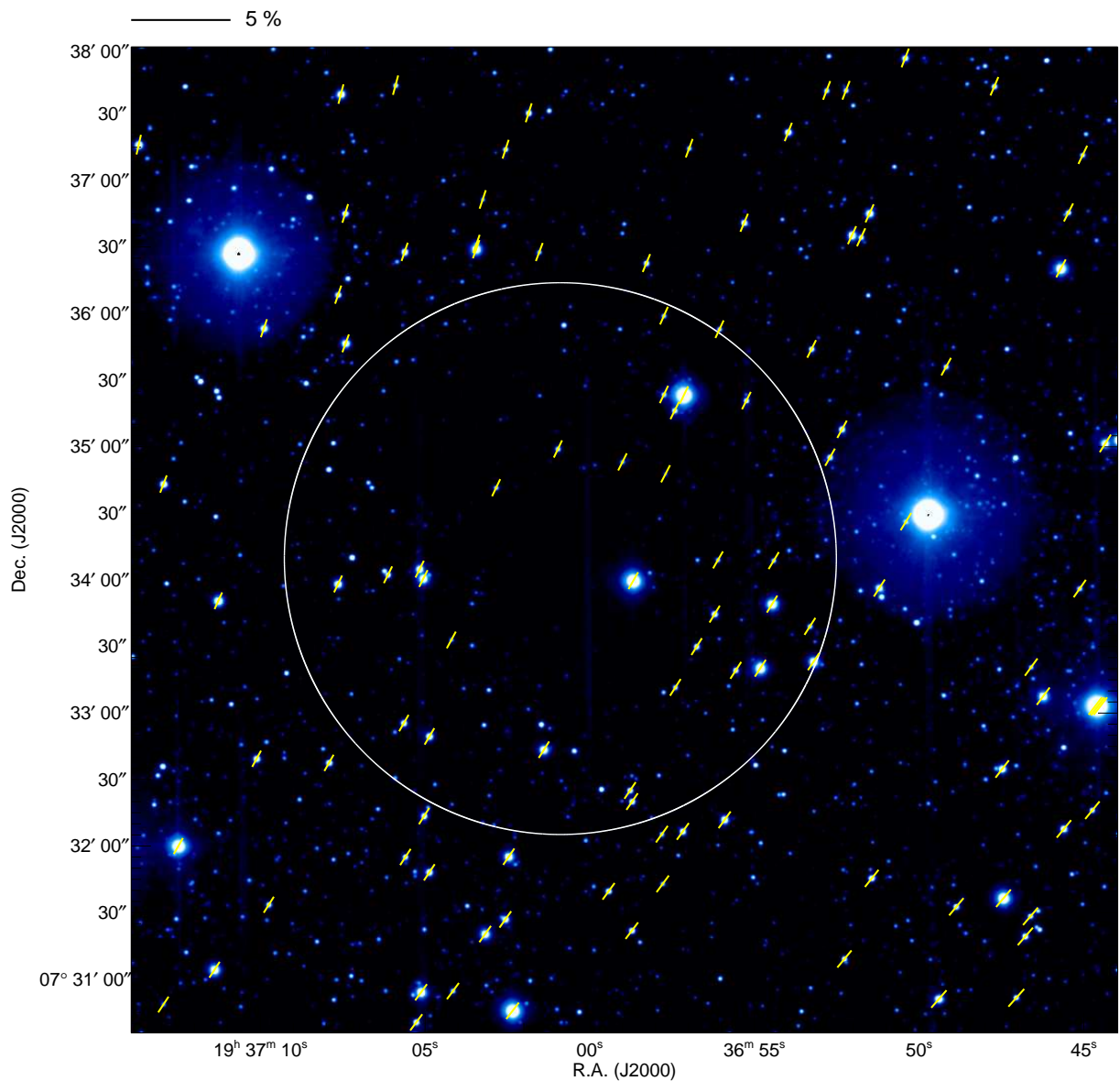}
\end{center}
 \caption{Estimated off-core polarization vectors superimposed on an $H$ band intensity image. Note that these vectors were not obtained directly from observations. Off-core vectors estimated by fitting are plotted at the position of each star. The white circle marks the core boundary (radius of 125$''$, Harvey et al. 2001). The scale of the 5\% polarization degree is shown above the image.}
   \label{fig1}
\end{figure}

\clearpage 
\begin{figure}[t]
\begin{center}
 \includegraphics[width=6.5 in]{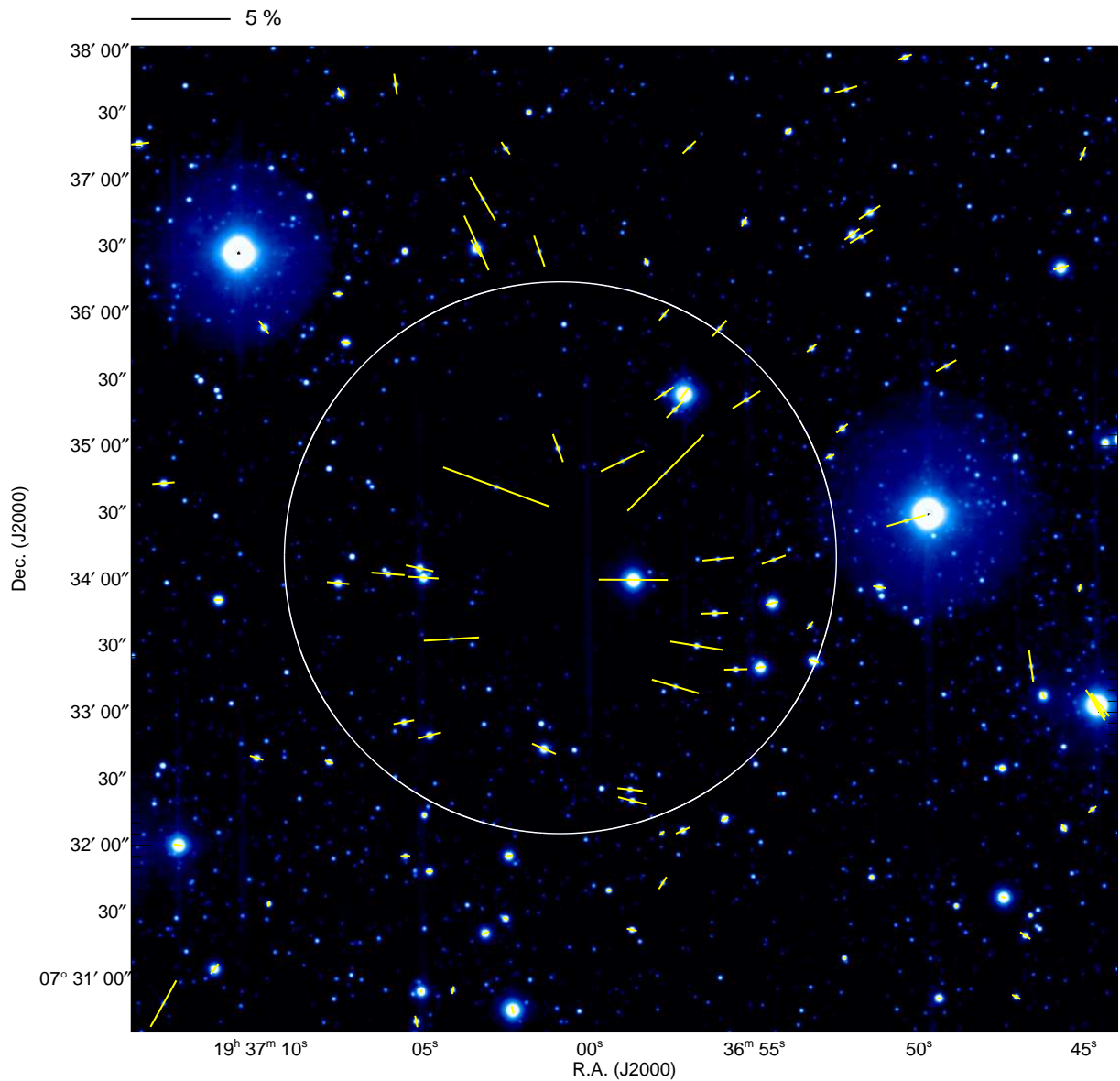}
\end{center}
 \caption{Polarization vectors after subtraction of the off-core component. The white circle marks the core boundary (radius of 125$''$, Harvey et al. 2001). The scale of the 5\% polarization degree is shown above the image.}
   \label{fig1}
\end{figure}

\clearpage 
\begin{figure}[t]
\begin{center}
 \includegraphics[width=6.5 in]{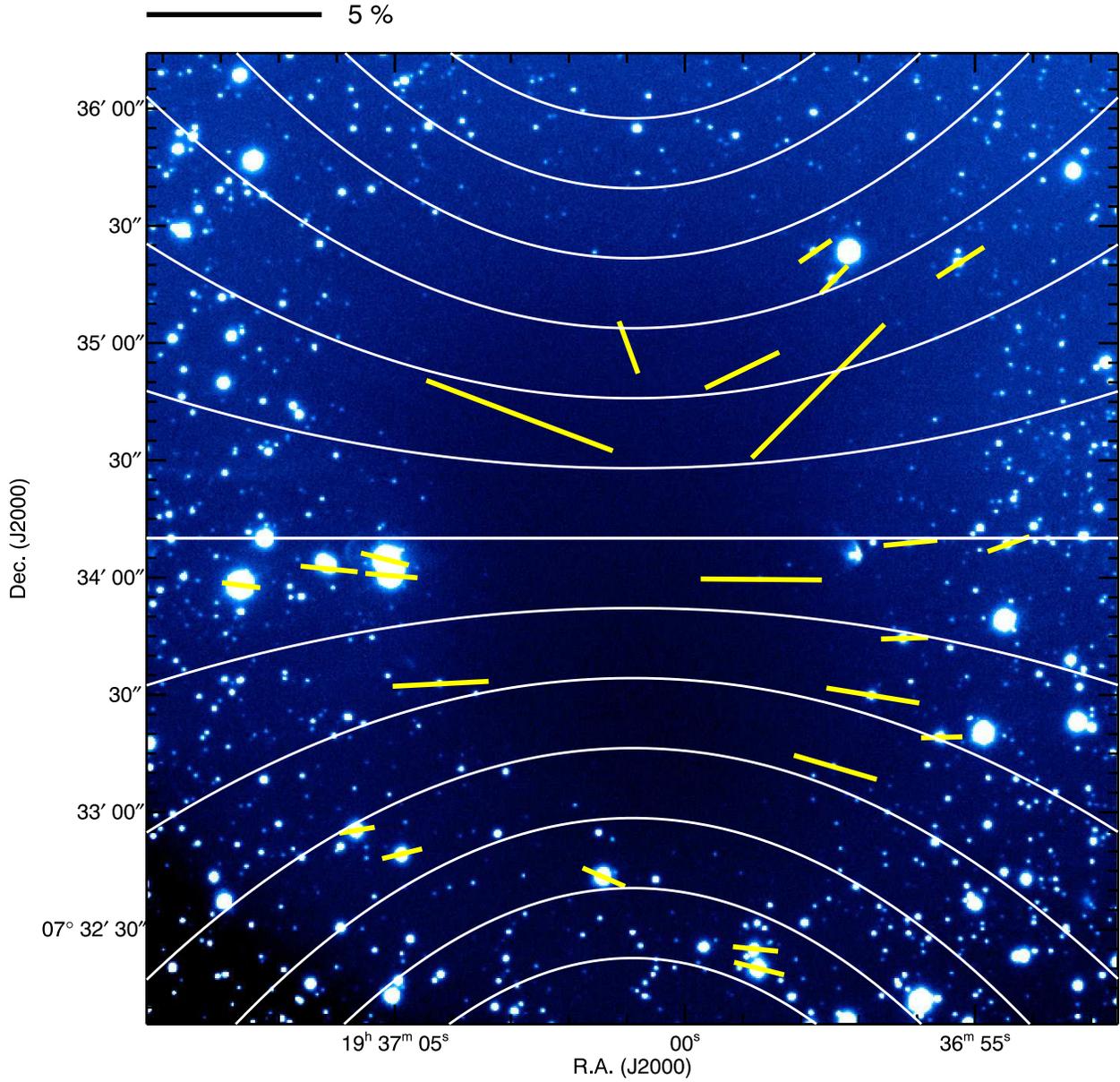}
\end{center}
 \caption{Polarization vectors after subtraction of the off-core component. The field of view is 250$''$ or 0.131 pc at a distance of 105 pc, which is equal to the diameter of the core. The background image is the optical (H$_{\alpha}$) image from Galf\aa lk \& Olofsson (2007), and the polarization vectors are from Figure 5. The white lines indicate the direction of the magnetic field inferred from parabolic fitting. The scale of the 5\% polarization degree is shown above the image.}
   \label{fig1}
\end{figure}

\clearpage 
\begin{figure}[t]
\begin{center}
 \includegraphics[width=6.5 in]{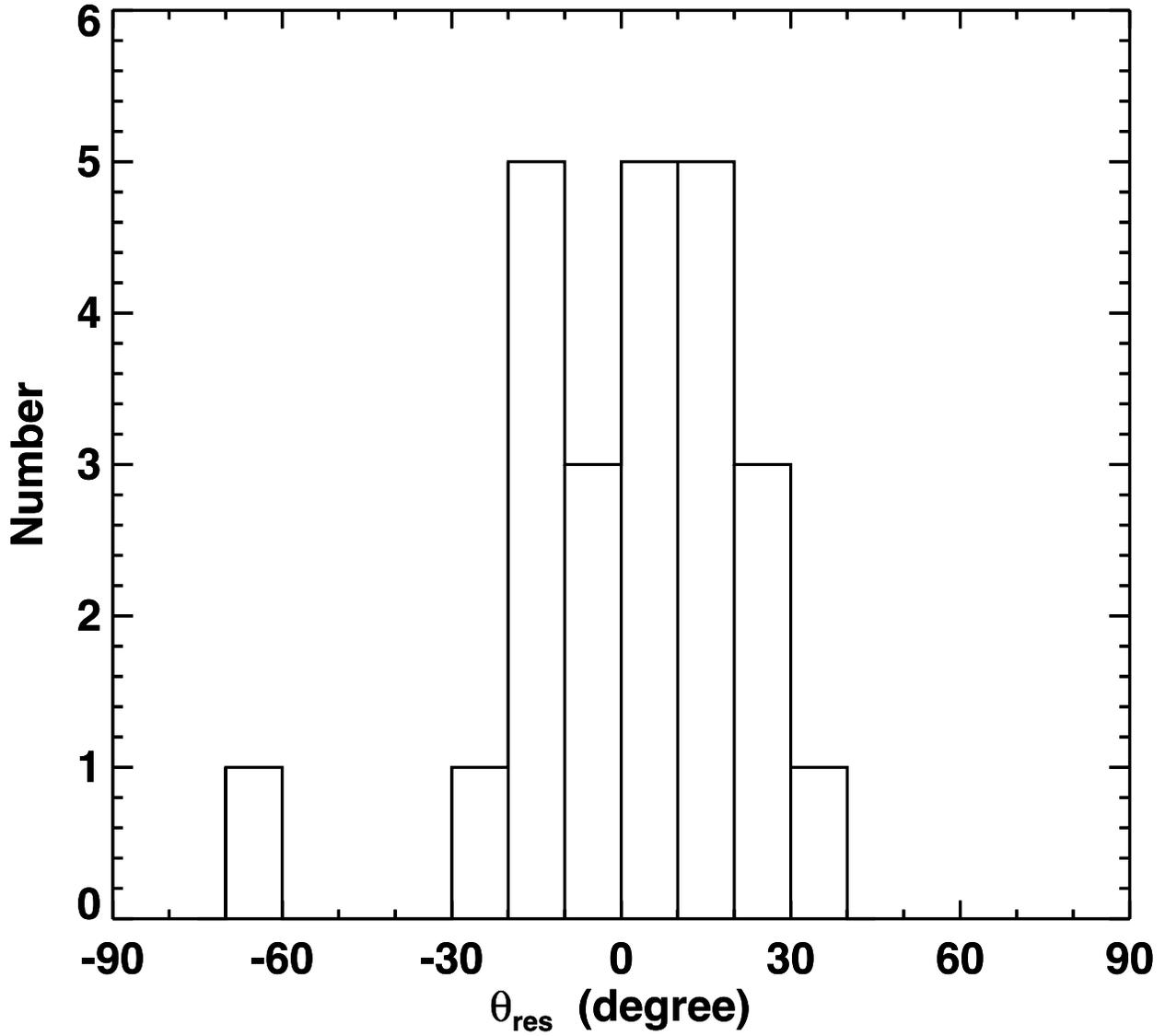}
\end{center}
 \caption{Histogram of the residual of the observed polarization angle after subtraction of the angle obtained by parabolic fitting ($\theta_{\rm res}$).}
   \label{fig1}
\end{figure}


\clearpage 
\begin{figure}[t]
\begin{center}
 \includegraphics[width=6.5 in]{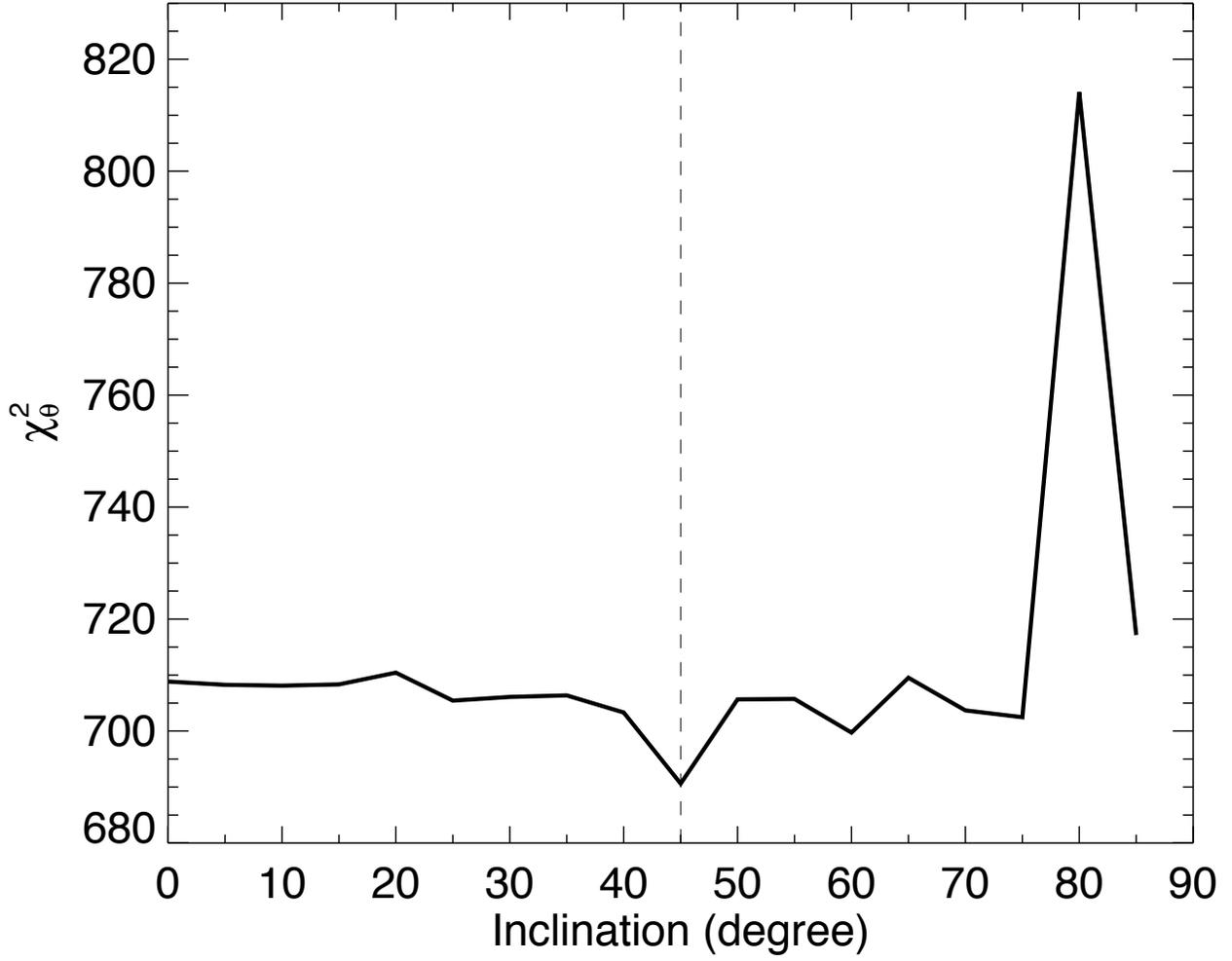}
\end{center}
 \caption{$\chi^{2}$ distribution for the polarization angle ($\chi^{2}_{\theta}$). The best magnetic curvature parameter ($C$) is determined at each $\gamma_{\rm mag}$. $\gamma_{\rm mag}=0^{\circ}$ and $90^{\circ}$ correspond to the edge-on and pole-on geometry in the magnetic axis.}
   \label{fig1}
\end{figure}

\clearpage 
\begin{figure}[t]
\begin{center}
 \includegraphics[width=6.5 in]{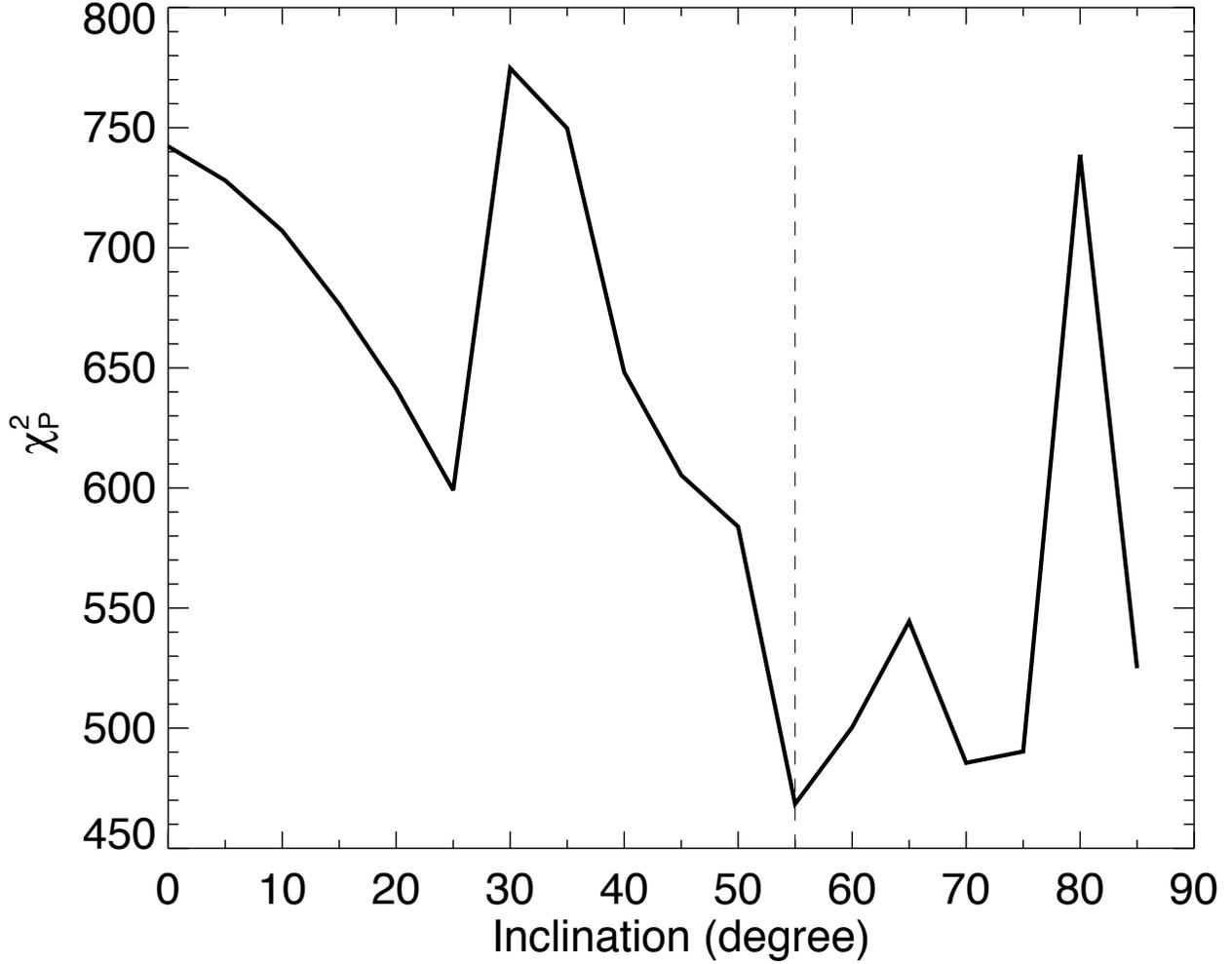}
\end{center}
 \caption{$\chi^{2}$ distribution for the polarization degree ($\chi^{2}_{P}$). $\gamma_{\rm mag}=0^{\circ}$ and $90^{\circ}$ correspond to the edge-on and pole-on geometry in the magnetic axis. Calculations of $\chi^2$ in polarization degree were performed after determining the best magnetic curvature parameter ($C$) which minimizes $\chi_2$ in the polarization angle. This calculation was carried out at each $\gamma_{\rm mag}$.}
   \label{fig1}
\end{figure}

\clearpage 
\begin{figure}[t]
\begin{center}
 \includegraphics[width=6.5 in]{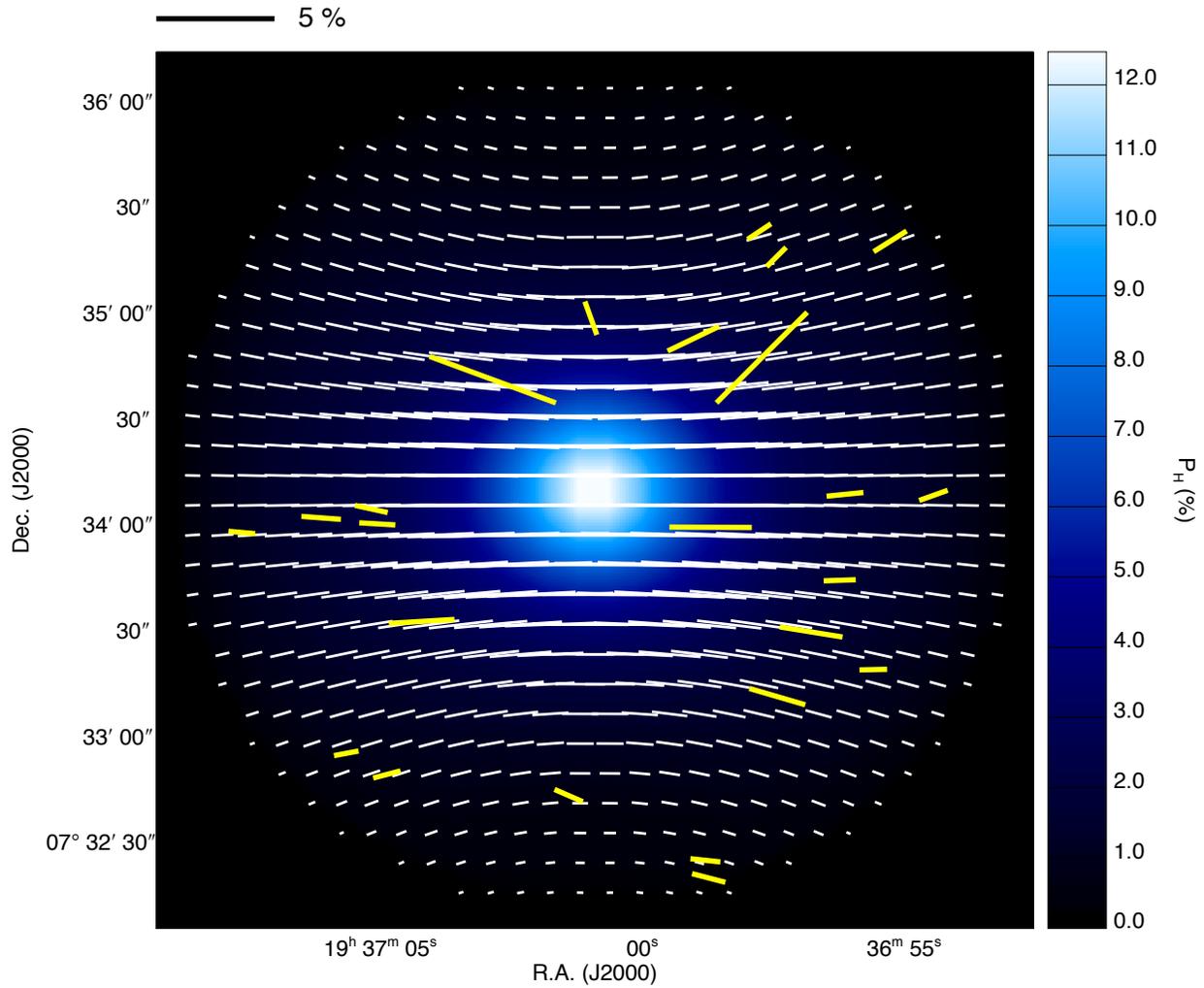}
\end{center}
 \caption{Best-fit 3D parabolic model (white vectors) with observed polarization vectors (yellow vectors). The background color image shows the polarization degree distribution of the best-fit model. The scale of the 5\% polarization degree is shown above the image.}
   \label{fig1}
\end{figure}

\clearpage 
\begin{figure}[t]
\begin{center}
 \includegraphics[width=6.5 in]{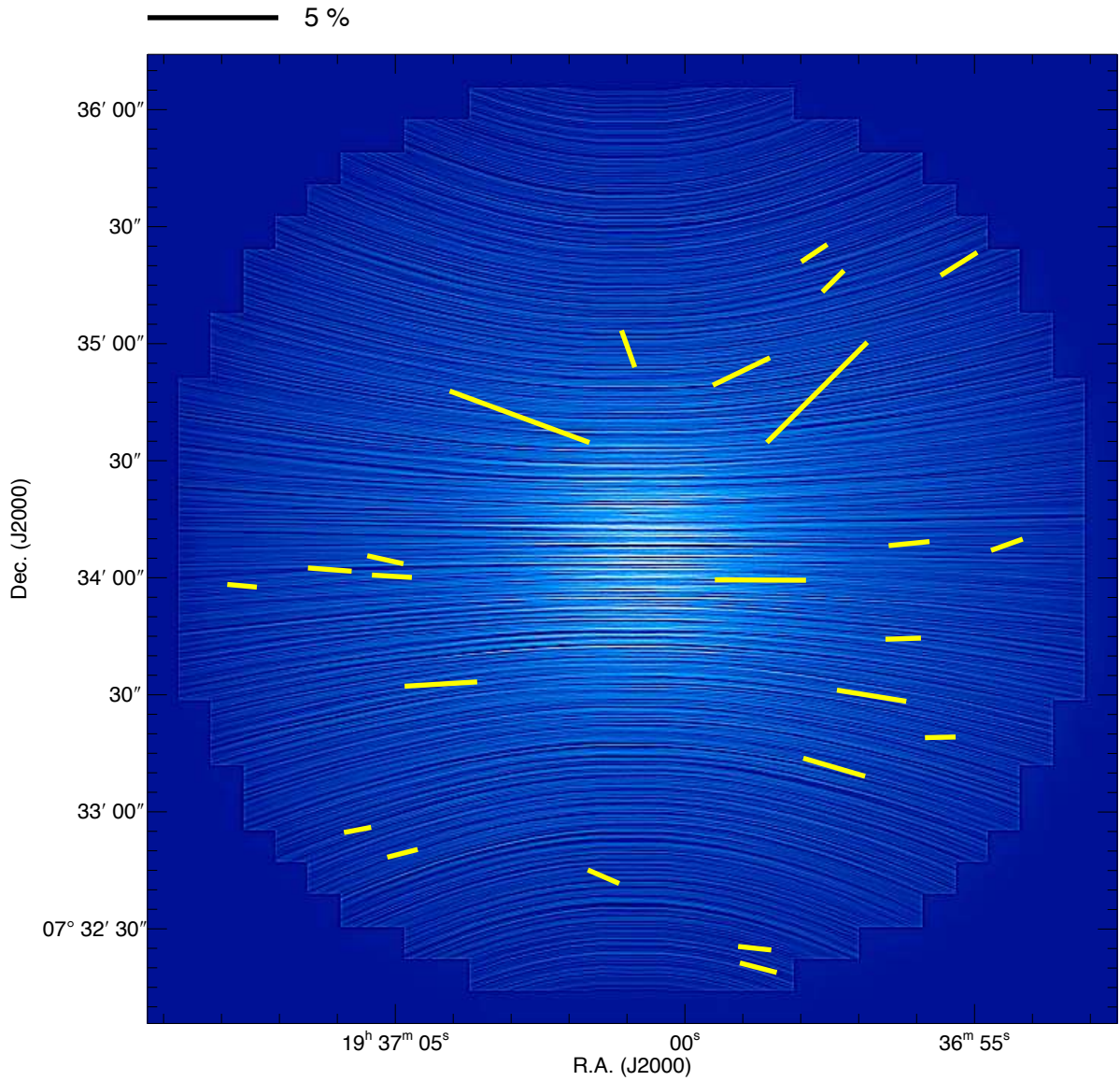}
\end{center}
 \caption{Same as Figure 10, but the background image was made using the line integral convolution (LIC) technique (Cabral \& Leedom 1993). The direction of the LIC \lq \lq texture'' is parallel to the direction of magnetic fields, and the background image is based on the polarization degree of model core.}
   \label{fig1}
\end{figure}

\clearpage 
\begin{figure}[t]  
\begin{center}
 \includegraphics[width=6.5 in]{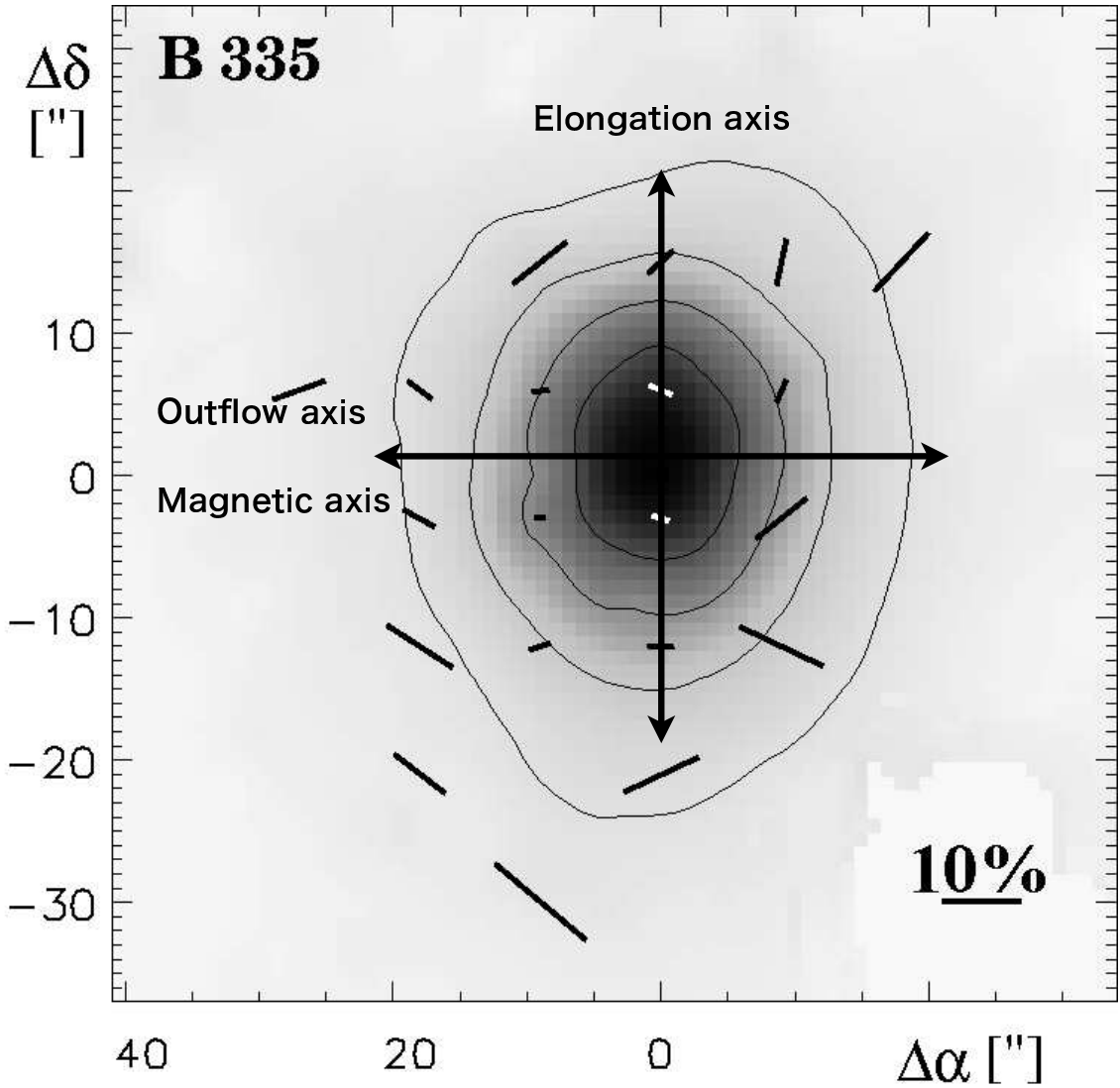}
\end{center}
 \caption{Relationship among the axis of elongation of the core ($\theta_{\rm elon} \sim 0^{\circ}$), the outflow axis ($\theta_{\rm rot} \sim 90^{\circ}$, Hirano et al. 1988), and the magnetic field axis ($\theta_{\rm mag}=90^{\circ}$) superimposed on the submm dust emission polarimetry map from Wolf et al. (2003).}
   \label{fig1}
\end{figure}

\clearpage 

\begin{figure}[t]  
\begin{center}
 \includegraphics[width=2.5 in]{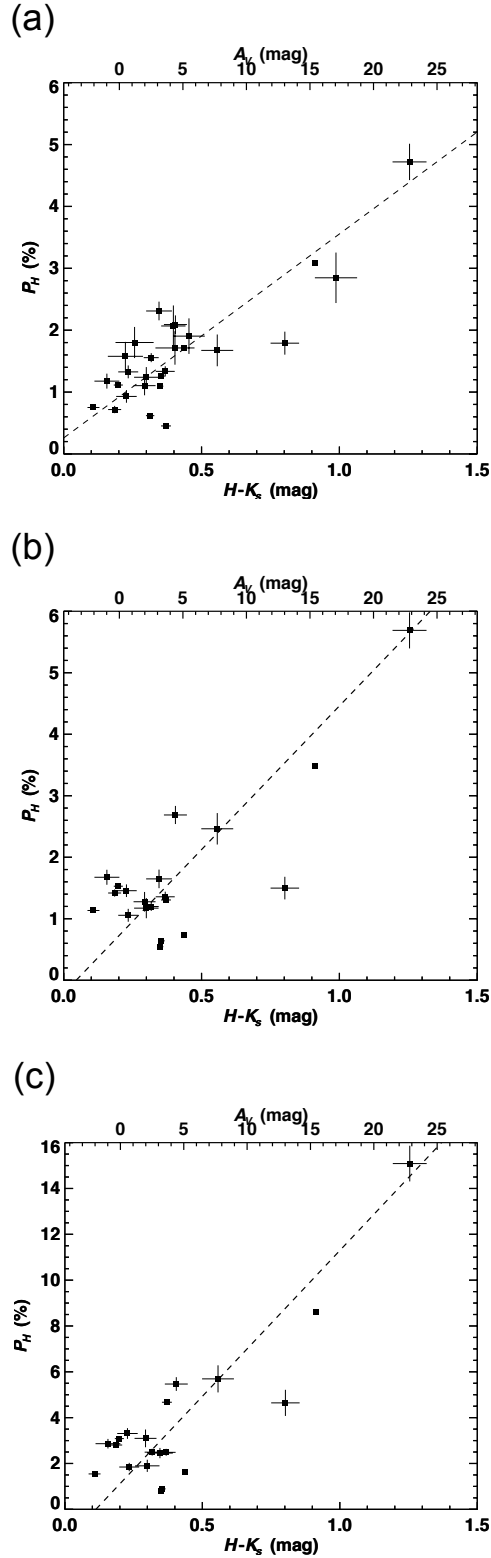}
\end{center}
 \caption{Relationship between polarization degree and $H-K_{\rm s}$ color toward background stars. The stars with $R \leq 125''$ and $P/\delta P \geq 6$ are plotted. In all the panels, the dashed lines denote the linear fit to the data. (a) $P$--$A$ relationship with no correction (original data). (b) $P$--$A$ relationship after correcting for ambient polarization components. (c) $P$--$A$ relationship after correcting for ambient polarization components, depolarization effect, and the magnetic inclination angle.}
   \label{fig1}
\end{figure}

\clearpage 

\begin{figure}[t]  
\begin{center}
 \includegraphics[width=6.5 in]{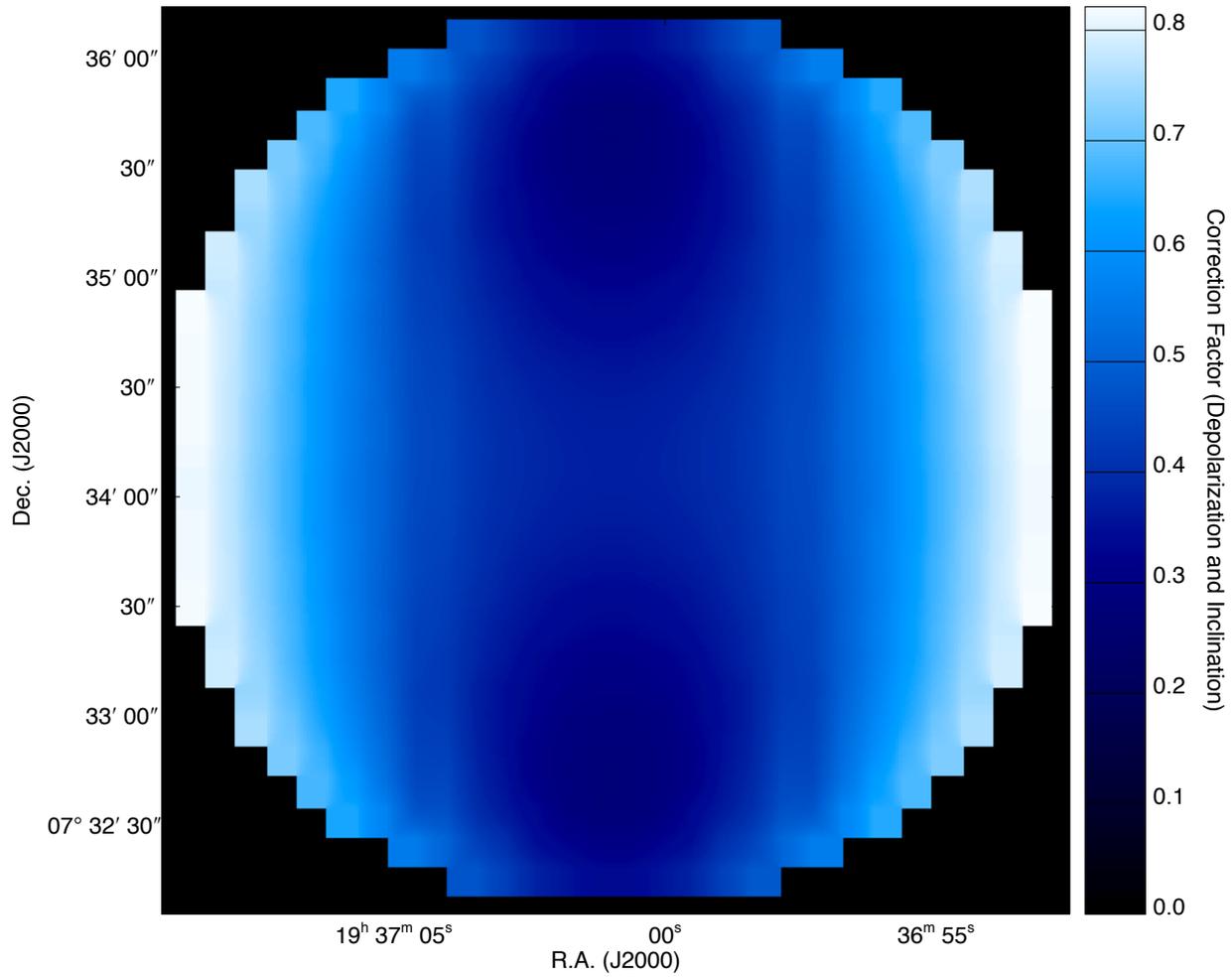}
\end{center}
 \caption{Distribution of the depolarization and inclination correction factors. The field of view is the same as the diameter of the core ($250''$).}
   \label{fig1}
\end{figure}

\clearpage 

\begin{figure}[t]  
\begin{center}
 \includegraphics[width=6.5 in]{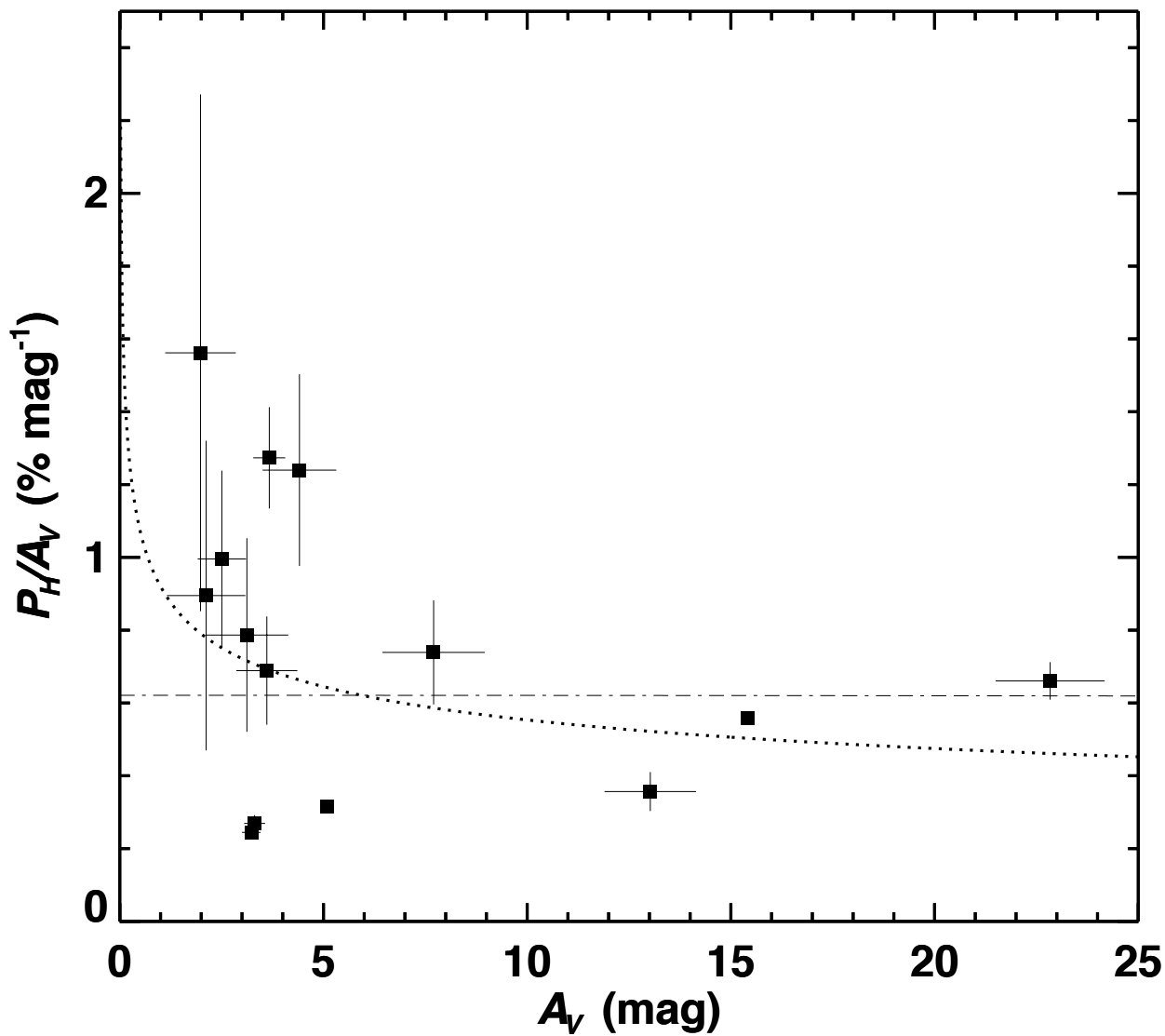}
\end{center}
 \caption{Relationship between polarization efficiency $P_H / A_V$ and $A_V$ toward background stars of FeSt 1-457. The stars with $R \leq 125''$ and $P/\delta P \geq 6$ are plotted. The dotted line shows the power-law fit to the data points. The dotted-dashed line shows the observational upper limit reported by Jones (1989).}
   \label{fig1}
\end{figure}

\end{document}